\documentclass[12pt]{article}
\begin{document}

\title{\bf \Large Symmetries of the Energy-Momentum Tensor of
Cylindrically Symmetric Static Spacetimes}

\author{M. Sharif \thanks{e-mail: hasharif@yahoo.com}
\\ Department of Mathematics, University of the Punjab,\\ Quaid-e-Azam
Campus Lahore-54590, PAKISTAN.}

\date{}

\maketitle

\begin{abstract}
We investigate matter symmetries of cylindrically symmetric static
spacetimes. These are classified for both cases when the
energy-momentum tensor is non-degenerate and also when it is
degenerate. It is found that the non-degenerate energy-momentum
tensor gives either {\it three}, {\it four}, {\it five}, {\it
six}, {\it seven} or {\it ten} independent matter collineations in
which {\it three} are isometries and the rest are proper. The
worth mentioning cases are those where we obtain the group of
matter collineations finite-dimensional even the energy-momentum
tensor is degenerate. These are either {\it three}, {\it four},
{\it five} or {\it ten}. Some examples are constructed satisfying
the constraints on the energy-momentum tensor.
\end{abstract}


\date{}

\newpage

\section{Introduction}

Let $M$ be a spacetime manifold with Lorentz metric $g$ of
signature $(+,-,-,-)$. It is assumed that the manifold $M$, and
the metric $g$, are smooth. There has been recent significant
interest in the study of the various symmetries (in particular,
Ricci and matter collineations) that arise in the exact solutions
of Einstein's field equations (EFEs)
\begin{equation}
R_{ab}-\frac{1}{2}Rg_{ab}\equiv G_{ab}=\kappa T_{ab}, \quad
(a,b=0,1,2,3),
\end{equation}
where $\kappa$ is the gravitational constant, $G_{ab}$ is the
Einstein tensor, $R_{ab}$ is the Ricci and $T_{ab}$ is the matter
(energy-momentum) tensor. Also, $R = g^{ab} R_{ab}$ is the Ricci
scalar. We have assumed here that the cosmological constant
$\Lambda=0$. The theoretical basis for the study of the affine,
conformal, projective, curvature (CCs) and Ricci collineations
(RCs) has been analysed and many examples have been discovered
[1]-[5]. The symmetries of the energy-momentum tensor have
recently been studied.

We define a differentiable vector field $\xi$ on $M$ to be a {\it
matter collineation} if
\begin{equation}
\pounds_\xi T_{ab}=0,
\end{equation}
where $\pounds$ is the Lie derivative operator, $\xi^a$ is the
symmetry or collineation vector. The study of matter collineations
(MCs) derives from the mathematical interest in the invariance
attributes of a geometrical object, i.e., Einstein tensor. Since
the Einstein tensor is related to the matter content of the
spacetime by the EFEs, the investigation of MCs seems to be more
relevant from the viewpoint of physics.

The study of symmetries played an important role in the
classification of spacetimes, giving rise to many interesting
results with useful applications. It is well known that two
different collineations are not in general equivalent. For
example, a Killing vector (KV) is a MC but the converse does not
hold. Collineations have been classified by means of their
relative properness by Katzin et al. [6,7]. This classification
indicates that the basic collineation is the KVs. The role of
isometries is to restric the general form of the metric.
Consequently, the number of independent field equations would
reduce and it would be easy to find the exact solutions. It is
noted that there are well known metrics which do not have
isometries [8]. This does not imply that they do not admit higher
symmetries. The symmetry properties given by KVs lead to
conservation laws [9]-[11]. A large number of solutions of the
EFEs with different symmetry structures have been found [10] and
classified according to their properties [12]. Symmetries of the
energy-momentum tensor (also called matter collineations) provide
conservation laws on matter fields. These enable us to know how
the physical fields, occupying in certain region of spacetimes,
reflect the symmetries of the metric [13].

There is a large body of recent literature which shows interest in
the study of MCs [14]-[23]. In a recent paper [22], the study of
MCs has been taken for spherically symmetric spacetimes and some
interesting results have been obtained. We have also classified
plane symmetric static spacetimes according to their MCs [23]. In
this paper, we extend the procedure to calculate MCs of
cylindrically symmetric static spacetimes both for non-degenerate
and also for degenerate cases. Here we would not give details of
the calculations as the procedure has been given in different
papers [22,23].

The MC Eq.(2) can be written in component form as
\begin{equation}
T_{ab,c} \xi^c + T_{ac} \xi^c_{,b} + T_{cb} \xi^c_{,a} = 0.
\end{equation}

The most general form of cylindrically symmetric static spacetime
is given by
\begin{equation}
ds^2=e^{\nu(r)}dt^2-dr^2-e^{\lambda(r)}d\theta^2-e^{\mu(r)}dz^2.
\end{equation}
The only non-zero components of the energy-momentum tensor, given
in Appendix A, are $T_{00},~T_{11},~T_{22},~T_{33}$. We can write
the MC equations as follows
\begin{equation}
T'_0\xi^1+2T_0\xi^0_{,0}=0,
\end{equation}
\begin{equation}
T'_1\xi^1+2T_1\xi^1_{,1}=0,
\end{equation}
\begin{equation}
T'_2\xi^1+2T_2\xi^2_{,2}=0,
\end{equation}
\begin{equation}
T'_3\xi^1+2T_3\xi^3_{,3}=0,
\end{equation}
\begin{equation}
T_0\xi^0_{0,1}+T_{1}\xi^1_{,0}=0,
\end{equation}
\begin{equation}
T_{0}\xi^0_{,2}+T_2\xi^2_{,0}=0,
\end{equation}
\begin{equation}
T_0\xi^0_{,3}+T_3\xi^3_{,0}=0,
\end{equation}
\begin{equation}
T_1\xi^1_{,2}+T_2\xi^2_{,1}=0,
\end{equation}
\begin{equation}
T_{1}\xi^1_{,3}+T_3\xi^3_{,1}=0,
\end{equation}
\begin{equation}
T_2\xi^2_{,3}+T_3\xi^3_{,2}=0,
\end{equation}
where prime $'$ indicates differentiation with respect to $r$.
These yield the first order non-linear coupled partial
differential equations in four variables $\xi^a(x^b)$. The
components of the energy-momentum tensor depend only on $r$. Here
we have used the notation $T_{aa}=T_a$ for the sake of brevity. We
solve this set of equations for the non-degenerate case, when
\begin{equation}
\det(T_{ab})=T_0T_1T_2T_3\neq 0
\end{equation}
and for the degenerate case, where $\det(T_a)=0$.

The rest of the paper is organized as follows. The next section
contains brief comments and results about MCs. In section 3, we
shall solve the MC equations when the energy-momentum tensor is
non-degenerate and in the next section MC equations are solved for
the degenerate energy-momentum tensor. In section 5, we shall
solve some of the constraints on energy-momentum tensor to obtain
exact solution of EFEs. Section 6 contains a summary and
discussion of the results obtained.

\section{Some General Comments}

Let $\xi$ be a matter collineation. All KVs, homothetic vectors
and special conformal Killing vectors are MCs. However, the
converse is not always true. In this case, the MC is called proper
or non-trivial. The study of MCs has many associated problems.
Here we list these in comparison with the other
symmetries.\\
\par \noindent
\par \noindent
1. When we define affine and conformal vector fields on $M$ we
usually assume that the vector field is at least $C^2$ and $C^3$
respectively. Then it follows from Hall et al. [4] that $\xi^a$
must be a smooth vector field on $M$. However, for $k\in Z^+$
there
exist MCs on smooth spacetimes which are $C^k$ but not $C^{k+1}$.\\
\par \noindent
\par \noindent
2. We know that an affine and conformal vector fields $\xi^a$ on
$M$ are uniquely determined by specifying $\xi^a$ and
$\xi^a{}_{;b}$ and respectively by specifying $\xi^a$ and the
components of its first two covariant derivatives $\xi^a{}_{;b}$
and $\xi^a{}_{;bc}$ at some point $p\in M$. However, the value of
$\xi^a$ and all its derivatives at some point $q\in M$ may not be
enough to determine uniquely a MC $\xi^a$ on $M$. The fact is that
two MCs which agree on a
non-empty open subset of $M$ may not agree on $M$.\\
\par \noindent
\par \noindent
3. The set of all MCs on $M$ is a vector space but in a similar
way to the sets of CCs and RCs, and unlike the sets of affine and
conformal vector fields, it could be infinite dimensional and
could fail to be a Lie algebra. The problem here arises from the
fact that such collineations must be $C^1$ in order that the
defining equations make sense. It is unfortunate that a matter,
Ricci or curvature collineation might turn out to be precisely
$C^1$ and so the differentiability may be destroyed under the Lie
bracket operation. On the other hand, if we assume that MCs are
$C^\infty$ then we recover the Lie algebra structure but we are
then forced to expel the collineations which are not smooth. The
infinite dimensionality may also lead to problems related to the
orbits of the resulting local diffeomorphism [4,24].\\
\par \noindent
\par \noindent
4. If the energy-momentum tensor is of rank 4 everywhere then we
can think of this tensor as a metric on the spacetime $M$. It then
follows from the theory of Killing vectors that the family of MCs
is, in fact, a Lie algebra of smooth vector fields on $M$, of
finite dimension, $\leq 10$, and, in addition, $\neq 9$ by
Fubini's theorem [12].

\section{Matter Collineations in the Non-Degenerate Case}

In this section, we shall evaluate MCs only for those cases which
have non-degenerate energy-momentum tensor, i.e., $T_a\neq 0$.

When we solve Eqs.(5)-(14) simultaneously, we get the following
constraint equation
\begin{equation}
T'_0\xi^1_{,23}=0.
\end{equation}
This equation implies that either
\begin{eqnarray*}
(1)\quad T'_0=0,\nonumber\\
or\quad (2)\quad T'_0\neq 0.
\end{eqnarray*}
{\bf Case 1:} In the first case we have $T_0=-k_1$, where $k_1$ is
a non-zero constant which implies that the MC equations yield the
following four possibilities:
\begin{eqnarray*}
(a_1)\quad T'_2=0,\quad T'_3=0,\nonumber\\
(a_2)\quad T'_2=0,\quad T'_3\neq 0,\nonumber\\
(a_3)\quad T'_2\neq 0,\quad T'_3=0,\nonumber\\
(a_4)\quad T'_2\neq 0,\quad T'_3\neq 0.
\end{eqnarray*}
{\bf Subcase 1$(a_1)$:} This implies that $T_2=k_2$ and $T_3=k_3$,
where $k_2$ and $k_3$ are non-zero constants. In this case, in
addition to the nonproper MCs $\xi_{(1)},~\xi_{(2)},~\xi_{(3)}$
given in Appendix B, we obtain the following proper MCs
\begin{eqnarray}
\xi_{(4)}&=&\theta\partial_t+\frac{k_1}{k_2}t
\partial_\theta,\nonumber\\
\xi_{(5)}&=&-\frac{1}{k_1}\int{\sqrt{T_1}}dr\partial_t
+\frac{1}{\sqrt{T_1}}t\partial_r,\nonumber\\
\xi_{(6)}&=&z\partial_t+\frac{k_1}{k_3}t
\partial_z,\nonumber\\
\xi_{(7)}&=&\frac{1}{\sqrt{T_1}}\partial_r,\nonumber\\
\xi_{(8)}&=&\frac{\theta}{\sqrt{T_1}}\partial_r
-\frac{1}{k_2}\int{\sqrt{T_1}}dr\partial_\theta,\nonumber\\
\xi_{(9)}&=&\frac{z}{\sqrt{T_1}}\partial_r
-\frac{1}{k_3}\int{\sqrt{T_1}}dr\partial_z,\nonumber\\
\xi_{(10)}&=&z\partial_\theta-\frac{k_2}{k_3}\theta
\partial_z.
\end{eqnarray}
Thus we have ten independent MCs in which seven are proper.\\
\par\noindent
{\bf Subcase 1$(a_2)$:} It follows that $T_2=k_2,~T_3'\neq 0$
which further yields the following two cases:
\begin{eqnarray*}
(b_1)\quad
[\frac{T_3}{\sqrt{T_1}}(\frac{T'_3}{2T_3\sqrt{T_1}})']'=0,\nonumber\\
(b_2)\quad
[\frac{T_3}{\sqrt{T_1}}(\frac{T'_3}{2T_3\sqrt{T_1}})']'\neq 0.
\end{eqnarray*}
The case 1$a_2(b_1)$ gives
$\frac{T_3}{\sqrt{T_1}}(\frac{T'_3}{2T_3\sqrt{T_1}})'=\alpha_1,$
where $\alpha_1$ is an arbitrary constant which may be
\begin{eqnarray*}
(c_1)\quad  \alpha_1>0,\quad (c_2)\quad  \alpha_1=0,\quad  or
\quad (c_3)\quad  \alpha_1<0.
\end{eqnarray*}
The first option 1$a_2b_1(c_1)$, when $\alpha_1>0$, gives the
following proper MCs
\begin{eqnarray}
\xi_{(4)}&=&\theta\partial_t +\frac{k_1}{k_2}t
\partial_\theta,\nonumber\\
\xi_{(5)}&=&\frac{e^{\sqrt{\alpha_1}z}}{\sqrt{T_1}}\partial_r
-\frac{T'_3e^{\sqrt{\alpha_1}z}}{2\sqrt{\alpha_1}T_3\sqrt{T_1}}
\partial_z,\nonumber\\
\xi_{(6)}&=&\frac{e^{-\sqrt{\alpha_1}z}}{\sqrt{T_1}}\partial_r
+\frac{T'_3e^{-\sqrt{\alpha_1}z}}{2\sqrt{\alpha_1}T_3\sqrt{T_1}}
\partial_z.
\end{eqnarray}
The second option 1$a_2b_1(c_2)$, when $\alpha_1=0$, yields the
following three proper MCs
\begin{eqnarray}
\xi_{(4)}&=&\theta\partial_t +\frac{k_1}{k_2}t
\partial_\theta,\nonumber\\
\xi_{(5)}&=&\frac{1}{T_1}\partial_r-\alpha_2z\partial_z,\nonumber\\
\xi_{(6)}&=&\frac{z}{\sqrt{T_1}}\partial_r
-(\int{\frac{\sqrt{T_1}}{T_3}}dr+\alpha_2\frac{z^2}{2})\partial_z,
\end{eqnarray}
where $\alpha_2=\frac{T'_3}{2T_3\sqrt{T_1}}$ is a non-zero
constant.

The third option 1$a_2b_1(c_3)$ gives the same MCs as the first
case 1$a_2b_1(c_1)$.

When we solve MC equations for the case 1$a_2(b_2)$, we have only
one proper MC given by
\begin{equation}
\xi_{(4)}=\theta\partial_t +\frac{k_1}{k_2}t\partial_\theta.
\end{equation}
{\bf Subcase 1$(a_3)$:} The subcase 1$(a_3)$ is similar to the the
subcase 1$(a_2)$ and MCs follow by interchanging $\theta$ and $z$
coordinates.\\
\par\noindent
{\bf Subcase 1$(a_4)$:} Here we have $T'_2\neq 0,~T'_3\neq 0$
which gives rise to the following two possibilities:
\begin{eqnarray*}
(b_1)\quad
[\frac{T_2}{\sqrt{T_1}}(\frac{T'_2}{2T_2\sqrt{T_1}})']'=0,\nonumber\\
(b_2)\quad
[\frac{T_2}{\sqrt{T_1}}(\frac{T'_2}{2T_2\sqrt{T_1}})']'\neq 0.
\end{eqnarray*}
The first possibility 1$a_4(b_1$) implies that
$\frac{T_2}{\sqrt{T_1}}(\frac{T'_2}{2T_2\sqrt{T_1}})'=\alpha_3$,
where $\alpha_3$ is an arbitrary constant such that
\begin{eqnarray*}
(c_1)\quad  \alpha_3>0,\quad (c_2)\quad  \alpha_3=0,\quad  or
\quad (c_3)\quad  \alpha_3<0.
\end{eqnarray*}
When $\alpha_3$ is positive, 1$a_4b_1(c_1)$, it further gives the
following two options:
\begin{eqnarray*}
(d_1)\quad
[\frac{T_3}{\sqrt{T_1}}(\frac{T'_3}{2T_3\sqrt{T_1}})']'=0,\nonumber\\
(d_2)\quad
[\frac{T_3}{\sqrt{T_1}}(\frac{T'_3}{2T_3\sqrt{T_1}})']'\neq 0.
\end{eqnarray*}
For the option 1$a_4b_1c_1(d_1)$, we obtain similar result to the
case 1$a_2(b_1)$. The option 1$a_4b_1c_1(d_2)$ gives the following
two cases according as
\begin{eqnarray*}
(e_1)\quad  \frac{T_2}{T_3}=constant\neq 0,\quad (e_2)\quad
\frac{T_2}{T_3}\neq constant.
\end{eqnarray*}
The case 1$a_4b_1c_1d_2(e_1)$ yields one proper MC
\begin{eqnarray}
\xi_{(4)}=z\partial_\theta-\frac{T_2}{T_3}\theta\partial_z,
\end{eqnarray}
and the case 1$a_4b_1c_1d_2(e_2)$ gives three independent MCs.

The case 1$a_4b_1(c_2)$, when $\alpha_3=0$, yields
$\frac{T'_2}{2T_2\sqrt{T_1}}=\alpha_4$, where $\alpha_4$ is a
non-zero constant which gives either
\begin{eqnarray*}
(d_1)\quad
[\frac{T_3}{\sqrt{T_1}}(\frac{T'_3}{2T_3\sqrt{T_1}})']'=0,\nonumber\\
or\quad (d_2)\quad
[\frac{T_3}{\sqrt{T_1}}(\frac{T'_3}{2T_3\sqrt{T_1}})']'\neq 0.
\end{eqnarray*}
The first possibility  1$a_4b_1c_2(d_1)$ further divides into
three cases according as
$\frac{T_3}{\sqrt{T_1}}(\frac{T'_3}{2T_3\sqrt{T_1}})'=\alpha_1$
\begin{eqnarray*}
(e_1)\quad  \alpha_1>0,\quad (e_2)\quad  \alpha_1=0,\quad  or
\quad (e_3)\quad  \alpha_1<0.
\end{eqnarray*}
The case 1$a_4b_1c_2d_1(e_1)$, when $\alpha_1>0$, MCs turn out to
be similar to the case 1$a_2b_1(c_1)$. When $\alpha_1=0$, i.e., in
the case 1$a_4b_1c_2d_1(e_2)$, we have further two options either
\begin{eqnarray*}
(f_1)\quad  \frac{T_3}{T_2}=k=constant\neq 0,\quad or \quad
(f_2)\quad \frac{T_3}{T_2}\neq constant.
\end{eqnarray*}
In the option 1$a_4b_1c_2d_1e_2(f_1)$, we obtain the following
four proper MCs
\begin{eqnarray}
\xi_{(4)}&=&-z\partial_\theta+\frac{\theta}{k}\partial_z,\nonumber\\
\xi_{(5)}&=&\frac{1}{\sqrt{T_1}}\partial_r
-\alpha_2\theta\partial_\theta-\alpha_2z\partial_z,\nonumber\\
\xi_{(6)}&=&\frac{\theta}{\sqrt{T_1}}\partial_r-\alpha_2\theta
z\partial_z-(\int\frac{\sqrt{T_1}}{T_2}dr+\alpha_2\frac{\theta^2}{2}
-k\alpha_2\frac{z^2}{2})\partial_\theta,\nonumber\\
\xi_{(7)}&=&\frac{z}{\sqrt{T_1}}\partial_r-\alpha_2\theta
z\partial_\theta-(\int\frac{\sqrt{T_1}}{T_2}dr-\alpha_2\frac{\theta^2}{2k}
+\alpha_2\frac{z^2}{2})\partial_z.
\end{eqnarray}
For the option 1$a_4b_1c_2d_1e_2(f_2)$, there is only one proper
MC
\begin{eqnarray}
\xi_{(4)}=\frac{1}{\sqrt{T_1}}\partial_r
+\theta\partial_\theta-\alpha_2z\partial_z.
\end{eqnarray}
The case 1$a_4b_1c_2d_1(e_3)$, when $\alpha_1<0$, yields similar
solution as the case 1$a_2b_1(c_3)$. The option 1$a_4b_1c_2(d_2)$
gives similar result as the case 1$a_4b_1c_1(d_2)$. The
possibility 1$a_4b_1(c_3)$, when $\alpha_3<0$, also gives similar
solution as the case 1$a_2b_1(c_1)$.

In the case 1$a_4(b_2)$, we have either
\begin{eqnarray*}
(c_1)\quad  \frac{T_2}{T_3}=constant,\quad or \quad (c_2)\quad
\frac{T_2}{T_3}\neq constant.
\end{eqnarray*}
For the case 1$a_4b_2(c_1)$, it coincides with 1$a_4b_1c_1(d_2)$
and in the case 1$a_4b_2(c_2)$, we get minimal MCs,
i.e., three.\\
\par\noindent
{\bf Case 2:} Now we evaluate MCs for the case when $T'_0\neq 0$.
This case implies that $\xi^1_{,23}=0$. Using this value in
Eqs.(12)-(14), we obtain
\begin{equation}
(\frac{T_2}{T_3})'\xi^2_{,3}=0.
\end{equation}
This implies that either
\begin{eqnarray*}
(a_1)\quad (\frac{T_2}{T_3})'=0,\quad or\quad (a_2)\quad
(\frac{T_2}{T_3})'\neq 0.
\end{eqnarray*}
{\bf Subcase 2$(a_1)$:} This yields $\frac{T_2}{T_3}=k_1$, where
$k_1$ is a non-zero constant. Solving MC equations using this
value, we obtain the following two cases:
\begin{eqnarray*}
(b_1)\quad (\frac{T'_2}{2T_2\sqrt{T_1}})'=0,\quad (b_2)\quad
(\frac{T'_2}{2T_2\sqrt{T_1}})'\neq 0.
\end{eqnarray*}
The case 2$a_1(b_1)$ implies that
$\frac{T'_2}{2T_2\sqrt{T_1}}=\beta_1$, a constant which yields
that either
\begin{eqnarray*}
(c_1)\quad \beta_1=0,\quad or\quad (c_2)\quad \beta_2\neq 0.
\end{eqnarray*}
For the case 2$a_1b_1(c_1)$, when $\beta_1=0$, it yields the
following two groups:
\begin{eqnarray*}
(d_1)\quad (\frac{(\sqrt{T_0})'}{\sqrt{T_1}})'=0,\quad (d_2)\quad
(\frac{(\sqrt{T_0})'}{\sqrt{T_1}})'\neq 0.
\end{eqnarray*}
The first group 2$a_1b_1c_1(d_1)$ gives ten independent MCs in
which seven are proper given by
\begin{eqnarray}
\xi_{(4)}&=&z\partial_\theta-k_1\theta\partial_z,\nonumber\\
\xi_{(5)}&=&\frac{1}{\sqrt{T_0}}\theta\sin k_2t\partial_t
-\frac{1}{\sqrt{T_1}}\theta\cos k_2t\partial_r
+\frac{\sqrt{T_0}}{k_2T_2}\cos k_2t\partial_\theta,\nonumber\\
\xi_{(6)}&=&-\frac{1}{\sqrt{T_0}}\theta\cos k_2t\partial_t
-\frac{1}{\sqrt{T_1}}\theta\sin k_2t\partial_r
+\frac{\sqrt{T_0}}{k_2T_2}\sin k_2t\partial_\theta,\nonumber\\
\xi_{(7)}&=&\frac{1}{k_1\sqrt{T_0}}z\sin k_2t\partial_t
-\frac{1}{k_1\sqrt{T_1}}z\cos k_2t\partial_r
+\frac{\sqrt{T_0}}{k_2T_2}\cos k_2t\partial_\theta,\nonumber\\
\xi_{(8)}&=&-\frac{1}{k_1\sqrt{T_0}}z\cos k_2t\partial_t
-\frac{1}{k_1\sqrt{T_1}}z\sin k_2t\partial_r
+\frac{\sqrt{T_0}}{k_2T_2}\sin k_2t\partial_\theta,\nonumber\\
\xi_{(9)}&=&\frac{1}{\sqrt{T_0}}\sin k_2t\partial_t
+\frac{1}{\sqrt{T_1}}\cos k_2t\partial_r,\nonumber\\
\xi_{(10)}&=&\frac{1}{\sqrt{T_0}}\cos k_2t\partial_t
+\frac{1}{\sqrt{T_1}}\sin k_2t\partial_r,
\end{eqnarray}
where $k_2=\frac{(\sqrt{T_0})'}{\sqrt{T_1}}$ is a non-zero
constant. The second group 2$a_1b_1c_1(d_2)$ gives further two
possibilities:
\begin{eqnarray*}
(e_1)\quad
\frac{T_0}{2\sqrt{T_1}}(\frac{T'_0}{T_0\sqrt{T_1}})'=constant=\beta_2,\nonumber\\
(e_2)\quad
\frac{T_0}{2\sqrt{T_1}}(\frac{T'_0}{T_0\sqrt{T_1}})'\neq constant.
\end{eqnarray*}
In the first possibility 2$a_1b_1c_1d_2(e_1)$, we further have two
options according as
\begin{eqnarray*}
(f_1)\quad \beta_2=0,\quad (f_2)\quad \beta_2\neq 0.
\end{eqnarray*}
The option  2$a_1b_1c_1d_2e_1(f_1)$ gives six independent MCs in
which three proper MCs are given by
\begin{eqnarray}
\xi_{(4)}&=&z\partial_\theta-k_1\theta\partial_z,\nonumber\\
\xi_{(5)}&=&(\frac{1}{\beta_3T_0}-\frac{\beta_3}{4}t^2)\partial_t
+\frac{1}{\sqrt{T_1}}t\partial_r,\nonumber\\
\xi_{(6)}&=&-\frac{\beta_3}{2}t\partial_t+\frac{1}{\sqrt{T_1}}\partial_r,
\end{eqnarray}
where $\beta_3=\frac{T'_0}{T_0\sqrt{T_1}}$ is a non-zero constant.
The possibility 2$a_1b_1c_1d_2e_1(f_2)$ also gives six independent
MCs. The three proper MCs are:
\begin{eqnarray}
\xi_{(4)}&=&z\partial_\theta-k_1\theta\partial_z,\nonumber\\
\xi_{(5)}&=&-\frac{T'_0}{2\sqrt{\beta_2}T_0\sqrt{T_1}}e^{\sqrt{\beta_2}t}
\partial_t+\frac{1}{\sqrt{T_1}}e^{\sqrt{\beta_2}t}\partial_r,\nonumber\\
\xi_{(6)}&=&\frac{T'_0}{2\sqrt{\beta_2}T_0\sqrt{T_1}}e^{-\sqrt{\beta_2}t}
\partial_t+\frac{1}{\sqrt{T_1}}e^{-\sqrt{\beta_2}t}\partial_r.
\end{eqnarray}
The case 2$a_1b_1c_1d_2(e_2)$ yields the same result as the case
1$a_4b_1c_1(d_2)$, i.e., one proper MC.

The possibility 2$a_1b_1(c_2)$, i.e., non-zero value of the
constant $\beta_1$, gives us two more options:
\begin{eqnarray*}
(d_1)\quad (\frac{T_2}{T_0})'=0,\quad (d_2)\quad
(\frac{T_2}{T_0})'\neq 0.
\end{eqnarray*}
The first option 2$a_1b_1c_2(d_1)$ gives ten independent MCs. The
seven proper MCs are
\begin{eqnarray}
\xi_{(4)}&=&z\partial_\theta-k_1\theta\partial_z,\nonumber\\
\xi_{(5)}&=&k_3\theta\partial_t+t\partial_\theta,\nonumber\\
\xi_{(6)}&=&\frac{k_3}{k_1}z\partial_t+t\partial_z,\nonumber\\
\xi_{(7)}&=&\frac{1}{2}(t^2-\frac{4}{\beta_1^2T_0}+k_3\theta^2
+\frac{k_3}{k_1}z^2)\partial_t-\frac{2}{\beta_1\sqrt{T_1}}t\partial_r
+t\theta\partial_\theta+tz\partial_z,\nonumber\\
\xi_{(8)}&=&-k_1t\theta\partial_t+\frac{2k_1}{\beta_1\sqrt{T_1}}
\theta\partial_r+\frac{1}{2}(-\frac{k_1}{k_3}t^2+\frac{4k_1}
{\beta_1^2T_2}-k_1\theta^2+z^2)\partial_\theta-k_1\theta z\partial_z,\nonumber\\
\xi_{(9)}&=&tz\partial_t-\frac{2}{\beta_1\sqrt{T_1}}z\partial_r+\theta
z\partial_\theta+\frac{1}{2}(\frac{k_1}{k_3}t^2
-\frac{4k_1}{\beta_1^2T_2}-k_1\theta^2+z^2)\partial_z,\nonumber\\
\xi_{(10)}&=&t\partial_t-\frac{2}{\beta_1\sqrt{T_1}}\partial_r+\theta
\partial_\theta+z\partial_z,
\end{eqnarray}
where $k_3=-\frac{T_2}{T_0}$ is a constant.

The second option 2$a_1b_1c_2(d_2)$ further yields two
possibilities whether
\begin{eqnarray*}
(e_1)\quad (\frac{T'_0}{T_0\sqrt{T_1}})'=0,\quad (e_2)\quad
(\frac{T'_0}{T_0\sqrt{T_1}})'\neq 0.
\end{eqnarray*}
If it is zero, i.e., the case 2$a_1b_1c_2d_2(e_1)$, we have the
following two proper MCs
\begin{eqnarray}
\xi_{(4)}&=&z\partial_\theta-k_1\theta\partial_z,\nonumber\\
\xi_{(5)}&=&\frac{\beta_4}{\beta_1}t\partial_t-\frac{2}
{\beta_1\sqrt{T_1}}\partial_r+\theta\partial_\theta+z\partial_z,
\end{eqnarray}
where $\beta_4=\frac{T'_0}{T_0\sqrt{T_1}}$ is a constant such that
$\beta_1\neq\beta_4$. If it is non-zero, i.e.,
2$a_1b_1c_2d_2(e_2)$, we have the same result as for the case
1$a_4b_1c_1(d_2)$.

The case 2$a_1(b_2)$, when $(\frac{T'_2}{2T_2\sqrt{T_1}})'\neq 0$,
implies that $T'_2\neq 0$ which results the following two
possibilities:
\begin{eqnarray*}
(c_1)\quad (\sqrt{\frac{T_0}{T_2}})'=0,\quad (c_2)\quad
(\sqrt{\frac{T_0}{T_2}})'\neq 0.
\end{eqnarray*}
The first possibility 2$a_1b_2(c_1)$ gives six independent MCs in
which three are the usual isometries and the remaining three are
proper MCs given by
\begin{eqnarray}
\xi_{(4)}&=&z\partial_\theta-k_1\theta\partial_z,\nonumber\\
\xi_{(5)}&=&\theta\partial_t+k_4^2t\partial_\theta,\nonumber\\
\xi_{(6)}&=&z\partial_t+k_1k_4^2t\partial_z,
\end{eqnarray}
where $k_4=\sqrt{\frac{T_0}{T_2}}$ is a constant. The second
possibility 2$a_1b_2(c_2)$ gives one proper MC as in the case
1$a_1b_1c_1(d_2)$.\\
\par\noindent
{\bf Subcase 2$(a_2)$:} Here we have $(\frac{T_2}{T_3})'\neq 0$.
If we use this constraint in MC Eqs.(7), (8) and (14), we have
\begin{equation}
T'_2\xi^1_{,3}=0=T'_3\xi^1_{,2}.
\end{equation}
This gives rise to the following three possibilities:
\begin{eqnarray*}
(b_1)\quad T'_2=0,\quad T'_3\neq 0,\nonumber\\
(b_2)\quad T'_2\neq 0,\quad T'_3=0,\nonumber\\
(b_3)\quad T'_2\neq 0,\quad T'_3\neq 0.
\end{eqnarray*}
In the first case, we can write
\begin{eqnarray}
\frac{T_0}{\sqrt{T_1}}(\frac{T'_0}{2T_0\sqrt{T_1}})'=\gamma_1,\quad
\frac{T_3}{\sqrt{T_1}}(\frac{T'_3}{2T_3\sqrt{T_1}})'=\gamma_2,
\end{eqnarray}
where $\gamma_1$ and $\gamma_2$ are arbitrary constants. From here
we have the following four different cases:
\begin{eqnarray*}
(c_1)\quad \gamma_1=0,\quad \gamma_2=0,\nonumber\\
(c_2)\quad \gamma_1\neq 0,\quad \gamma_2=0,\nonumber\\
(c_3)\quad \gamma_1=0,\quad \gamma_2\neq 0,\nonumber\\
(c_4)\quad \gamma_1\neq 0,\quad \gamma_2\neq 0.
\end{eqnarray*}
The first case 2$a_2b_1(c_1)$ can be divided into the following
two options according as
\begin{eqnarray*}
(d_1)\quad (\frac{T_0}{T_3})'=0,\quad or \quad (d_2)\quad
(\frac{T_0}{T_3})'\neq 0.
\end{eqnarray*}
The first option 2$a_2b_1c_1(d_1)$ gives seven independent MCs in
which four are proper
\begin{eqnarray}
\xi_{(4)}&=&(\frac{\gamma_3}{2}t^2-\int{\frac{\sqrt{T_1}}{T_0}}dr
+\frac{\gamma_3}{2k_1}z^2)\partial_t+\frac{1}{\sqrt{T_1}}t\partial_r
+\gamma_3tz\partial_z,\nonumber\\
\xi_{(5)}&=&\gamma_3tz\partial_t+\frac{1}{\sqrt{T_1}}z\partial_r
+(\frac{k_1\gamma_3}{2}t^2-\int{\frac{\sqrt{T_1}}{T_3}}dr
+\frac{\gamma_3}{2}z^2)\partial_z,\nonumber\\
\xi_{(6)}&=&\gamma_3t\partial_t+\frac{1}{\sqrt{T_1}}\partial_r
+\gamma_3z\partial_z,\nonumber\\
\xi_{(7)}&=&z\partial_t+k_1t\partial_z,
\end{eqnarray}
where $k_1=-\frac{T_0}{T_3}$ and
$\gamma_3=-\frac{T'_0}{2T_0\sqrt{T_1}}$ are constants. The second
option 2$a_2b_1c_1(d_2)$ yields only one proper MC given by
\begin{equation}
\xi_{(4)}=\gamma_3t\partial_t+\frac{1}{\sqrt{T_1}}t\partial_r
+\gamma_4tz\partial_z,
\end{equation}
where $\gamma_4=-\frac{T'_3}{2T_3\sqrt{T_1}}$.

In the case 2$a_2b_1(c_2)$, when $\gamma_1\neq 0,\quad
\gamma_2=0$, we have either
\begin{eqnarray*}
(d_1)\quad \gamma_1>0,\quad or\quad (d_2)\quad \gamma_1<0.
\end{eqnarray*}
For 2$a_2b_1c_2(d_1)$, when $\gamma_1>0$, we obtain three
independent MCs which are the usual isometries.

The case 2$a_2b_1c_2(d_2)$, when $\gamma_1<0$, we have two options
\begin{eqnarray*}
(e_1)\quad (\frac{T_0}{T_3})'=0,\quad or\quad (e_2)\quad
(\frac{T_0}{T_3})'\neq 0.
\end{eqnarray*}
For the case 2$a_2b_1c_2d_2(e_1)$, we have only one proper MC
given by
\begin{equation}
\xi_{(4)}=\frac{1}{k_1}z\partial_t+t\partial_z.
\end{equation}
In the case 2$a_2b_1c_2d_2(e_2)$, we obtain the minimal symmetry.

The case 2$a_2b_1(c_3)$, when $\gamma_1=0,\quad \gamma_2\neq 0$,
is similar to the previous case 2$a_2b_1(c_2)$ by interchanging
$t$ and $z$.

The case 2$a_2b_1(c_4)$, when $\gamma_1\neq 0,\quad \gamma_2\neq
0$, yields the following four different possibilities:
\begin{eqnarray*}
(d_1)\quad \gamma_1>0,\quad \gamma_2>0,\nonumber\\
(d_2)\quad \gamma_1>0,\quad \gamma_2<0,\nonumber\\
(d_3)\quad \gamma_1<0,\quad \gamma_2>0,\nonumber\\
(d_4)\quad \gamma_1<0,\quad \gamma_2<0.
\end{eqnarray*}
The first possibility further gives two options according as
\begin{eqnarray*}
(e_1)\quad \gamma_2
T_0\int{\frac{\sqrt{T_1}}{T_0}}dr+\frac{T'_3}{2\sqrt{T_1}}=0,\nonumber\\
or \quad (e_2)\quad \gamma_2
T_0\int{\frac{\sqrt{T_1}}{T_0}}dr+\frac{T'_3}{2\sqrt{T_1}}\neq 0.
\end{eqnarray*}
The first option 2$a_2b_1c_4d_1(e_1)$ gives the minimal symmetry.
The second option 2$a_2b_1c_4d_1(e_2)$ further gives two
possibilities
\begin{eqnarray*}
(f_1)\quad (\frac{T_0}{T_3})'=0,\quad (f_2)\quad
(\frac{T_0}{T_3})'\neq 0.
\end{eqnarray*}
The case 2$a_2b_1c_4d_1e_2(f_1)$ yields four independent MCs in
which the proper MC is given by
\begin{equation}
\xi_{(4)}=z\partial_t+k_1t\partial_z.
\end{equation}
For the case 2$a_2b_1c_4d_1e_2(f_2)$, we have three independent
MCs.

All other cases 2$a_2b_1c_4(d_2-d_4)$ are similar to the previous
case 2$a_2b_1c_4(d_1)$.

The case 2$a_2(b_2)$, when $T'_2\neq 0,\quad T'_3=0$, is similar
to the first case 2$a_2(b_1)$.

In the third case 2$a_2(b_3)$, when $T'_2\neq 0,\quad T'_3\neq 0$,
we further have the following two possibilities:
\begin{eqnarray*}
(c_1)\quad (\frac{T'_2}{2T_2\sqrt{T_1}})'=0,\quad
(\frac{T'_3}{2T_3\sqrt{T_1}})'=0,\nonumber\\
(c_2)\quad (\frac{T'_2}{2T_2\sqrt{T_1}})'\neq 0,\quad
(\frac{T'_3}{2T_3\sqrt{T_1}})'\neq 0.
\end{eqnarray*}
For the first possibility 2$a_2b_3(c_1)$, we obtain the following
three options:
\begin{eqnarray*}
(d_1)\quad (\frac{T_2}{T_0})'=0,\quad
(\frac{T_3}{T_0})'\neq 0,\nonumber\\
(d_2)\quad (\frac{T_2}{T_0})'\neq 0,\quad
(\frac{T_3}{T_0})'=0,\nonumber\\
(d_3)\quad (\frac{T_2}{T_0})'\neq 0,\quad (\frac{T_3}{T_0})'\neq
0.
\end{eqnarray*}
The first option 2$a_2b_3c_1(d_1)$ yields five independent MCs in
which three are the usual KVs and the remaining are the proper MCs
given by
\begin{eqnarray}
\xi_{(4)}&=&\gamma_3t\partial_t+\frac{1}{\sqrt{T_1}}\partial_r
+\gamma_3\theta\partial_\theta+\gamma_4z\partial_z,\nonumber\\
\xi_{(5)}&=&k_2\theta\partial_t+t\partial_\theta,
\end{eqnarray}
where $k_2=-\frac{T_2}{T_0}$ is a non-zero constant. The second
option 2$a_2b_3c_1(d_2)$ is similar to the first one. The third
case 2$a_2b_3c_1(d_3)$ implies that either
\begin{eqnarray*}
(e_1)\quad (\frac{T'_0}{2T_0\sqrt{T_1}})'=0,\quad or\quad
(e_2)\quad (\frac{T'_0}{2T_0\sqrt{T_1}})'\neq 0.
\end{eqnarray*}
For the first option 2$a_2b_3c_1d_3(e_1)$, we get one proper MC
given by
\begin{equation}
\xi_{(4)}=\gamma_3t\partial_t+\frac{1}{\sqrt{T_1}}\partial_r
+\gamma_3\theta\partial_\theta+\gamma_4z\partial_z,
\end{equation}
In the second option 2$a_2b_3c_1d_3(e_2)$, we obtain the minimal
symmetry.

The case 2$a_2b_3(c_2)$ also yields the same three possibilities:
\begin{eqnarray*}
(d_1)\quad (\frac{T_2}{T_0})'=0,\quad
(\frac{T_3}{T_0})'\neq 0,\nonumber\\
(d_2)\quad (\frac{T_2}{T_0})'\neq 0,\quad
(\frac{T_3}{T_0})'=0,\nonumber\\
(d_3)\quad (\frac{T_2}{T_0})'\neq 0,\quad (\frac{T_3}{T_0})'\neq
0.
\end{eqnarray*}
It is to be noted that we have excluded the possibility when both
are constants as this leads $\frac{T_2}{T_3}$ to be constant which
gives a contradiction. The first case 2$a_2b_3c_2(d_1)$ gives only
one proper MC, i.e.
\begin{equation}
\xi_{(4)}=k_2\theta\partial_t+t\partial_\theta.
\end{equation}
The second case 2$a_2b_3c_2(d_2)$ is similar to the previous one
and the third case 2$a_2b_3c_2(d_3)$ gives the minimal MCs.

\section{Matter Collineations in the Degenerate Case}

In this section only those cases will be considered for which
$det(T_{ab})=0$ which implies that at least one of the components
of the energy-momentum tensor is zero, i.e., $T_a=0$. The trivial
case is that when all $T_a$ are zero. In this case, every
direction in a MC.
The remaining cases can be divided into three main groups:\\
1. When at least one of $T_a$ is non-zero;\\
2. When at least two of $T_a$ are non-zero;\\
3. When three of $T_a$ are non-zero.\\
\par\noindent
{\bf Case 1:} This case can further be divided into the following
four subcases:
\begin{eqnarray*}
(a_1)\quad T_0=0,\quad T_1=0,\quad T_2=0,\quad T_3\neq 0,\nonumber\\
(a_2)\quad T_0=0,\quad T_1=0,\quad T_2\neq 0,\quad T_3=0,\nonumber\\
(a_3)\quad T_0=0,\quad T_1\neq 0,\quad T_2=0,\quad T_3=0,\nonumber\\
(a_4)\quad T_0\neq 0,\quad T_1=0,\quad T_2=0,\quad T_3=0.
\end{eqnarray*}
When we use the values of 1$(a_1)$ in MC equations, we obtain
$\xi^3=\xi^3(z)$ and Eq.(9) gives
\begin{equation}
\xi^1=-\frac{2T_3}{T'_3}\xi^3_{,3}(z),
\end{equation}
where $T'_3\neq 0$ and $\xi^0,~\xi^1$ are arbitrary functions of
$t,r,\theta,z$. This gives infinite dimensional MCs. The second
case 1$(a_2)$ is similar to the first one if we interchange the
indices $2$ and $3$.

The third case 1$(a_3)$ gives $\xi^1=\xi^1(r)$ and Eq.(7) yields
\begin{equation}
\xi^1=\frac{c_1}{\sqrt{T_1}},
\end{equation}
where $\xi^0,~\xi^2,~\xi^3$ are arbitrary functions of
$t,r,\theta,z$ which gives infinite dimensional MCs. The fourth
case 1$(a_4)$ yields the similar result as the case 1$(a_1)$ by
interchanging the indices $0$ and $3$. Thus we obtain infinite
dimensional MCs in all the possibilities of the case 1.\\
\par\noindent
{\bf Case 2:} This case has the following six possibilities:
\begin{eqnarray*}
(a_1)\quad T_0=0,\quad T_1=0,\quad T_2\neq 0,\quad T_3\neq 0,\nonumber\\
(a_2)\quad T_0=0,\quad T_1\neq 0,\quad T_2=0,\quad T_3\neq 0,\nonumber\\
(a_3)\quad T_0=0,\quad T_1\neq 0,\quad T_2\neq 0,\quad T_3=0,\nonumber\\
(a_4)\quad T_0\neq 0,\quad T_1=0,\quad T_2=0,\quad T_3\neq 0,\nonumber\\
(a_5)\quad T_0\neq 0,\quad T_1=0,\quad T_2\neq 0,\quad T_3=0,\nonumber\\
(a_6)\quad T_0\neq 0,\quad T_1\neq 0,\quad T_2=0,\quad T_3=0.
\end{eqnarray*}
When we replace the information of the subcase 2$(a_1)$ in MC
equations, we obtain
$\xi^0=\xi^0(t,r,\theta,z),~\xi^2=\xi^2(\theta,z),~
\xi^3=\xi^3(\theta,z)$ and
\begin{equation}
(\frac{T_3}{T_2})'\xi^3_{,2}=0.
\end{equation}
From here we have two options:
\begin{eqnarray*}
(b_1)\quad (\frac{T_3}{T_2})'=0,\quad (b_2)\quad
(\frac{T_3}{T_2})'\neq 0.
\end{eqnarray*}
For the first option 2$a_1(b_1)$, MC equations yield
\begin{equation}
\xi^2_{,22}+\frac{1}{c}\xi^2_{,33}=0,
\end{equation}
\begin{equation}
\xi^3_{,22}+\frac{1}{c}\xi^3_{,33}=0,
\end{equation}
where $c=\frac{T_3}{T_2}$ is a non-zero constant. If $c>0$,
Eqs.(43) and (44) yield the following solution
\begin{equation}
\xi^2=f_+(\theta+\frac{\iota z}{\sqrt{c}})+f_-(\theta-\frac{\iota
z}{\sqrt{c}}),
\end{equation}
\begin{equation}
\xi^3=g_+(\theta+\frac{\iota z}{\sqrt{c}})+g_-(\theta-\frac{\iota
z}{\sqrt{c}}).
\end{equation}
Replacing the value of $\xi^2$ in Eq.(8), we obtain
\begin{equation}
\xi^1=-\frac{2T_2}{T'_2}[f_{+,2}(\theta+\frac{\iota
z}{\sqrt{c}})+f_{-,2}(\theta-\frac{\iota z}{\sqrt{c}})],
\end{equation}
where $T'_2\neq 0$ and $\xi^0=\xi^0(t,r,\theta,z)$. If we take
$T_2=constant$, MC equations give the following solution
\begin{eqnarray}
\xi^0&=&\xi^0(t,r,\theta,z),\quad
\xi^1=\xi^1(t,r,\theta,z),\nonumber\\
\xi^2&=&c_1z+c_2,\quad \xi^3=c_1\theta+c_3.
\end{eqnarray}
The option 2$a_1(b_2)$ further divides into three cases:
\begin{eqnarray*}
(c_1)\quad T'_2=0,\quad T'_3\neq 0,\nonumber\\
(c_2)\quad T'_2\neq 0,\quad T'_3=0,\nonumber\\
(c_3)\quad T'_2\neq 0,\quad T'_3\neq 0.
\end{eqnarray*}
In the case 2$a_1b_2(c_1)$, we get the following solution
\begin{eqnarray}
\xi^0&=&\xi^0(t,r,\theta,z),\quad
\xi^1=\frac{T_3}{T'_3}f(z),\nonumber\\
\xi^2&=&c_1,\quad \xi^3=-\frac{1}{2}\int{f(z)dz}+c_2.
\end{eqnarray}
The case 2$a_1b_2(c_2)$ gives the similar results as the case
2$a_1b_2(c_1)$ by interchanging $\theta$ and $z$.

If we solve MC equations for the possibility 2$a_1b_2(c_3)$, we
have the following solution
\begin{eqnarray}
\xi^0&=&\xi^0(t,r,\theta,z),\quad
\xi^1=\frac{T_2}{T'_2}c_1,\nonumber\\
\xi^2&=&c_1\theta+c_2,\quad \xi^3=c_3z+c_4.
\end{eqnarray}
For the subcase 2$(a_2)$, we further have the following two
options from MC equations
\begin{eqnarray*}
(b_1)\quad
[\frac{T_3}{\sqrt{T_1}}(\frac{T'_3}{2T_3\sqrt{T_1}})']'=0,\nonumber\\
(b_2)\quad
[\frac{T_3}{\sqrt{T_1}}(\frac{T'_3}{2T_3\sqrt{T_1}})']'\neq 0.
\end{eqnarray*}
Also $\xi^0=\xi^0(t,r,\theta,z),\quad \xi^1=\xi^1(r,z),\quad
\xi^2=\xi^2(r),\quad \xi^3=\xi^3(r,z)$. The first possibility
2$a_2(b_1)$ gives
$\frac{T_3}{\sqrt{T_1}}(\frac{T'_3}{2T_3\sqrt{T_1}})'=c$, where
$c$ is an arbitrary constant which implies that either
\begin{eqnarray*}
(c_1)\quad c>0,\quad (c_2)\quad c=0,\quad or \quad c<0.
\end{eqnarray*}
The case 2$a_2b_1(c_1)$ yields the following solution
\begin{eqnarray}
\xi^0&=&\xi^0(t,r,\theta,z),\nonumber\\
\xi^1&=&\frac{1}{\sqrt{T_1}}(c_1e^{\sqrt{c}z}+c_2e^{-\sqrt{c}z}),
\quad \xi^2=\xi^2(r,z),\nonumber\\
\xi^3&=&-\frac{T'_3}{2\sqrt{c}T_3\sqrt{T_1}}(c_1e^{\sqrt{c}z}
-c_2e^{-\sqrt{c}z})+c_3.
\end{eqnarray}
For the case 2$a_2b_1(c_2)$, when $c=0$, we obtain
\begin{eqnarray}
\xi^0&=&\xi^0(t,r,\theta,z),\nonumber\\
\xi^1&=&\frac{1}{\sqrt{T_1}}(c_1z+c_2),\quad
\xi^2=\xi^2(r,z)\nonumber\\
\xi^3&=&-c_1\int{\frac{\sqrt{T_1}}{T_3}}dr
-\frac{T'_3}{2T_3\sqrt{T_1}}(c_1\frac{z^2}{2}+c_2z)+c_3.
\end{eqnarray}
In the case 2$a_2b_1(c_3)$, when $c<0$, the solution is similar to
the previous case 2$a_2b_1(c_2)$.

The second possibility 2$a_2(b_2)$ gives
\begin{eqnarray}
\xi^0&=&\xi^0(t,r,\theta,z),\quad \xi^1=0,\nonumber\\
\xi^2&=&\xi^2(r,z),\quad \xi^3=c_1.
\end{eqnarray}
The subcase 2$(a_3)$ gives similar results as the case 2$(a_2)$ by
interchanging the indices $2$ and $3$.

The subcases 2$(a_4)$ and 2$(a_5)$ yield results similar to the
case 2$(a_1)$ if we interchange indices $0$, $2$ and $0$, $3$
respectively.

The subcase 2$(a_6)$ would give similar result as the case
2$(a_2)$ by interchanging the indices $0$, $3$. It is to be
noted that we again have infinite dimensional MCs in the case 2.\\
\par\noindent
{\bf Case 3:} The case, when only one component of the
energy-momentum tensor is zero, can have four different subcases:
\begin{eqnarray*}
(a_1)\quad T_0=0,\quad T_1\neq 0,\quad T_2\neq 0,\quad T_3\neq 0,\nonumber\\
(a_2)\quad T_0\neq 0,\quad T_1=0,\quad T_2\neq 0,\quad T_3\neq 0,\nonumber\\
(a_3)\quad T_0\neq 0,\quad T_1\neq 0,\quad T_2=0,\quad T_3\neq 0,\nonumber\\
(a_4)\quad T_0\neq 0,\quad T_1\neq 0,\quad T_2\neq 0,\quad T_3=0.
\end{eqnarray*}
The first subcase gives $\xi^0=\xi^0(t,r,\theta,z)$ with the
following two possibilities:
\begin{eqnarray*}
(b_1)\quad [\frac{T_2}{\sqrt{T_1}}(-\frac{T'_2}
{2T_2\sqrt{T_1}})']'=0,\nonumber\\
(b_2)\quad [\frac{T_2}{\sqrt{T_1}}(-\frac{T'_2}
{2T_2\sqrt{T_1}})']'\neq 0.
\end{eqnarray*}
In the first possibility 3$a_1(b_1)$, we have
$\frac{T_2}{\sqrt{T_1}}(-\frac{T'_2} {2T_2\sqrt{T_1}})'=\alpha_1$,
where $\alpha_1$ is an arbitrary constant which can be such that
\begin{eqnarray*}
(c_1)\quad \alpha_1>0,\quad (c_2)\quad \alpha_1=0,\quad (c_3)\quad
\alpha_1<0 .
\end{eqnarray*}
The case 3$a_1b_1(c_1)$, when $\alpha_1>0$ gives further three
options according as $\alpha_2=\frac{T_3}{\sqrt{T_1}}(-\frac{T'_3}
{2T_3\sqrt{T_1}})'$
\begin{eqnarray*}
(d_1)\quad \alpha_2>0,\quad (d_2)\quad \alpha_2=0,\quad (d_3)\quad
\alpha_2<0 .
\end{eqnarray*}
The first option 3$a_1b_1c_1(d_1)$, when $\alpha_2>0$, gives
either
\begin{eqnarray*}
(e_1)\quad
\sqrt{\frac{\alpha_1}{\alpha_2}}+\sqrt{\frac{\alpha_2}{\alpha_1}}=0,\quad
or\quad (e_2)\quad
\sqrt{\frac{\alpha_1}{\alpha_2}}+\sqrt{\frac{\alpha_2}{\alpha_1}}\neq
0.
\end{eqnarray*}
For the first case 3$a_1b_1c_1d_1(e_1)$, we obtain
\begin{eqnarray}
\xi^0&=&\xi^0(t,r,\theta,z),\quad \xi^1=0,\nonumber\\
\xi^2&=&c_1z+c_2,\quad \xi^3=c_1\theta+c_3.
\end{eqnarray}
In the second case 3$a_1b_1c_1d_1(e_2)$, we have the solution
\begin{eqnarray}
\xi^0&=&\xi^0(t,r,\theta,z),\quad \xi^1=0,\nonumber\\
\xi^2&=&c_1z+c_2,\quad \xi^3=-c_1k_1\theta+c_3,
\end{eqnarray}
where $k_1=\frac{T_2}{T_3}$ is a constant.

The second option 3$a_1b_1c_1(d_2)$, when $\alpha_2=0$, yields the
following solution
\begin{eqnarray}
\xi^0&=&\xi^0(t,r,\theta,z),\nonumber\\
\xi^1&=&\frac{1}{\sqrt{T_1}}[e^{\iota\sqrt{\alpha_1}\theta}(c_1z+c_2)
+e^{-\iota\sqrt{\alpha_1}\theta}(c_3z+c_4)],\nonumber\\
\xi^2&=&-\frac{\iota T'_2\sqrt{T_1}}{2T_2\sqrt{T_1}}
[e^{\iota\sqrt{\alpha_1}\theta}(c_1z+c_2)
-e^{-\iota\sqrt{\alpha_1}\theta}(c_3z+c_4)]+c_5,\nonumber\\
\xi^3&=&-\alpha_3[e^{\iota\sqrt{\alpha_1}\theta}(c_1\frac{z^2}{2}+c_2z)
+e^{-\iota\sqrt{\alpha_1}\theta}(c_3\frac{z^2}{2}+c_4z)]\nonumber\\
&-&[c_1e^{\iota\sqrt{\alpha_1}\theta}+c_3e^{-\iota\sqrt{\alpha_1}\theta}]
\int{\frac{\sqrt{T_1}}{T_3}dr+c_6},
\end{eqnarray}
where $\alpha_3=\frac{T'_3}{2T_3\sqrt{T_1}}$ is a constant.

The third option 3$a_1b_1c_1(d_3)$, for $\alpha_2<0$, is similar
to the first option 3$a_1b_1c_1(d_1)$.

Now we come to the case 3$a_1b_1(c_2)$ when $\alpha_1=0$ which
gives $\alpha_4=-\frac{T'_2}{2T_2\sqrt{T_1}}$, a constant such
that
\begin{eqnarray*}
(d_1)\quad \alpha_4=0,\quad (d_2)\quad \alpha_4\neq 0.
\end{eqnarray*}
In the case 3$a_1b_1c_2(d_1)$, we have
$\alpha_2=\frac{T_3}{\sqrt{T_1}}(-\frac{T'_3}{2T_3\sqrt{T_1}})'$
which yields the following options:
\begin{eqnarray*}
(e_1)\quad \alpha_2>0,\quad (e_2)\quad \alpha_2=0,\quad (e_3)\quad
\alpha_2<0.
\end{eqnarray*}
For the case 3$a_1b_1c_2d_1(e_1)$, when $\alpha_2>0$, we have the
following MCs
\begin{eqnarray}
\xi^0&=&\xi^0(t,r,\theta,z),\nonumber\\
\xi^1&=&\frac{1}{\sqrt{T_1}}[e^{\iota\sqrt{\alpha_2}\theta}(c_1z+c_2)
+e^{-\iota\sqrt{\alpha_2}\theta}(c_3z+c_4)],\nonumber\\
\xi^2&=&0,\nonumber\\
\xi^3&=&-\frac{T'_3}{\iota2\sqrt{\alpha_2}T_3\sqrt{T_1}}
(c_1e^{\iota\sqrt{\alpha_2}z}-c_2e^{-\iota\sqrt{\alpha_2}z}).
\end{eqnarray}
The case 3$a_1b_1c_2d_1(e_2)$, when $\alpha_2=0$, we have further
two possibilities according as
$\alpha_5=-\frac{T'_3}{2T_3\sqrt{T_1}}$, a constant such that
\begin{eqnarray*}
(f_1)\quad \alpha_5=0,\quad (f_2)\quad \alpha_5\neq 0.
\end{eqnarray*}
For the case 3$a_1b_1c_2d_1e_2(f_1)$, we get the following result
\begin{eqnarray}
\xi^0&=&\xi^0(t,r,\theta,z),\quad
\xi^1=\frac{1}{\sqrt{T_1}}(c_1\theta+c_2z+c_3),\nonumber\\
\xi^2&=&-\frac{c_1}{T_2}\int{\sqrt{T_1}}dr,\quad
\xi^3=-\frac{c_2}{T_3}\int{\sqrt{T_1}}dr.
\end{eqnarray}
For $\alpha_5$ to be non-zero, i.e., the case
3$a_1b_1c_2d_1e_2(f_2)$, we obtain
\begin{eqnarray}
\xi^0&=&\xi^0(t,r,\theta,z),\quad
\xi^1=\frac{1}{\sqrt{T_1}}(c_1z+c_2),\nonumber\\
\xi^2&=&0,\quad
\xi^3=\alpha_5(c_1\frac{z^2}{2}+c_2z)-c_1\int{\frac{\sqrt{T_1}}{T_3}}dr.
\end{eqnarray}
The case 3$a_1b_1c_2d_1(e_3)$ when $\alpha_2<0$ is similar to the
case 3$a_1b_1c_2d_1(e_1)$.

In the case 3$a_1b_1c_2(d_2)$ when $\alpha_2\neq 0$, we have
$\alpha_2=\frac{T_3}{\sqrt{T_1}}(-\frac{T'_3}{2T_3\sqrt{T_1}})'$,
a constant which gives the following three options:
\begin{eqnarray*}
(e_1)\quad \alpha_2>0,\quad (e_2)\quad \alpha_2=0,\quad (e_3)\quad
\alpha_2<0.
\end{eqnarray*}
The case 3$a_1b_1c_2d_2(e_1)$ yields the following solution
\begin{eqnarray}
\xi^0&=&\xi^0(t,r,\theta,z),\quad
\xi^1=0,\nonumber\\
\xi^2&=&(c_1z+c_2),\quad \xi^3=-\frac{T_2}{T_3}c_1\theta+c_3.
\end{eqnarray}
The case 3$a_1b_1c_2d_2(e_2)$, when $\alpha_2=0$, gives the
following MCs
\begin{eqnarray}
\xi^0&=&\xi^0(t,r,\theta,z),\quad
\xi^1=\frac{c_1}{T_1},\nonumber\\
\xi^2&=&(\alpha_3c_1\theta+c_2),\quad \xi^3=\alpha_3c_1z+c_3.
\end{eqnarray}
The last possibility 3$a_1b_1c_2d_2(e_3)$ when $\alpha_2$ is
negative yields similar solution to the positive case
3$a_1b_1c_2d_2(e_1)$.

The case 3$a_1b_1(c_3)$, when $\alpha_1<0$, is similar to the case
3$a_1b_1(c_1)$.

The case 3$a_1(b_2)$ yields the following solution
\begin{eqnarray}
\xi^0&=&\xi^0(t,r,\theta,z),\quad \xi^1=0,\nonumber\\
\xi^2&=&c_1z+c_2,\quad \xi^3=c_3\theta+c_4.
\end{eqnarray}

The case 3$(a_2)$ when $T_0\neq 0,\quad T_1=0,\quad T_2\neq
0,\quad T_3\neq 0$ gives the following two options:
\begin{eqnarray*}
(b_1)\quad T'_0=0,\quad (b_2)\quad T'_0\neq 0.
\end{eqnarray*}
The first option 3$a_2(b_1)$ gives either
\begin{eqnarray*}
(c_1)\quad (\frac{T'_2T_3}{T_2T'_3})'=0,\quad or\quad (c_2)\quad
(\frac{T'_2T_3}{T_2T'_3})'\neq 0.
\end{eqnarray*}
The case 3$a_2b_1(c_1)$ gives one proper MC
\begin{equation}
\xi_{(4)}=-\frac{2T_2}{T'_2}\beta_1\partial_r
+\beta_1\theta\partial_\theta+z\partial_z,
\end{equation}
where $\beta_1=\frac{T'_2T_3}{T_2T'_3}$. For the case
3$a_2b_1(c_2)$, we have either
\begin{eqnarray*}
(d_1)\quad T'_2=0=T'_3,\quad or\quad (d_2)\quad T'_2\neq 0,\quad
T'_3\neq 0.
\end{eqnarray*}
The case 3$a_2b_1c_2(d_1)$ yields infinite dimensional MCs. For
the case 3$a_2b_1c_2(d_2)$, we get three MCs which are the usual
KVs.

The case 3$a_2(b_2)$ divides into four groups:
\begin{eqnarray*}
(c_1)\quad (\frac{T_0}{T_2})'=0,\quad
(\frac{T_0}{T_3})'=0, \nonumber\\
(c_2)\quad (\frac{T_0}{T_2})'=0,\quad
(\frac{T_0}{T_3})'\neq 0, \nonumber\\
(c_3)\quad (\frac{T_0}{T_2})'\neq 0,\quad
(\frac{T_0}{T_3})'=0, \nonumber\\
(c_4)\quad (\frac{T_0}{T_2})'\neq 0,\quad (\frac{T_0}{T_3})'\neq
0.
\end{eqnarray*}
The first group 3$a_2b_2(c_1)$ gives ten independent MCs in which
seven are proper MCs given by
\begin{eqnarray}
\xi_{(4)}&=&-\frac{k_1}{k_2}z\partial_\theta+\theta\partial_z,\nonumber\\
\xi_{(5)}&=&\theta\partial_t+k_1t\partial_\theta,\nonumber\\
\xi_{(6)}&=&z\partial_t+k_2t\partial_z,\nonumber\\
\xi_{(7)}&=&(\frac{\theta^2}{2}+\frac{k_1z^2}{2k_2}
+k_1\frac{t^2}{2})\partial_t-\frac{2k_1T_0}{T'_0}t\partial_r
+k_1t\theta\partial_\theta+k_1tz\partial_z,\nonumber\\
\xi_{(8)}&=&t\theta\partial_t-\frac{2T_0}{T'_0}\theta\partial_r
+(k_1\frac{t^2}{2}+\frac{\theta^2}{2}-\frac{k_1}{k_2}\frac{z^2}{2})
\partial_\theta+\theta z\partial_z,\nonumber\\
\xi_{(9)}&=&tz\partial_t-\frac{2T_0}{T'_0}z\partial_r+\theta
z\partial_\theta+(k_2\frac{t^2}{2}-\frac{k_2}{k_1}
\frac{\theta^2}{2}+\frac{z^2}{2})\partial_z,\nonumber\\
\xi_{(10)}&=&t\partial_t-\frac{2T_0}{T'_0}\partial_r+\theta
\partial_\theta+z\partial_z,
\end{eqnarray}
where $k_1=-\frac{T_0}{T_2}$ and $k_2=-\frac{T_0}{T_3}$ are
non-zero constants.

The second group 3$a_2b_2(c_2)$ when $(\frac{T_0}{T_2})'$=0,\quad
$(\frac{T_0}{T_3})'\neq 0$ can give two more possibilities whether
\begin{eqnarray*}
(d_1)\quad (\frac{T'_3T_0}{T_3T'_0})'=constant,\quad (d_2)\quad
(\frac{T'_3T_0}{T_3T'_0})'\neq constant
\end{eqnarray*}
The case 3$a_2b_2c_2(d_1)$, we have two proper MC given by
\begin{eqnarray}
\xi_{(4)}&=&\theta\partial_t+k_1t\partial_\theta,\nonumber\\
\xi_{(5)}&=&t\partial_t-\frac{2T_0}{T'_0}\partial_r
+\theta\partial_\theta+cz\partial_z,
\end{eqnarray}
where $\frac{T_0}{T_3}=c$ is an arbitrary constant. If it is not
constant, i.e., the case 3$a_2b_2c_2(d_2)$, we obtain only one
proper MC given by
\begin{equation}
\xi_{(4)}=\theta\partial_t+k_1t\partial_\theta.
\end{equation}

The third group 3$a_2b_2(c_3)$ when $(\frac{T_0}{T_2})'\neq
0$,\quad $(\frac{T_0}{T_3})'=0$ would give the similar solution as
the previous one 3$a_2b_2(c_2)$.

The last group 3$a_2b_2(c_4)$ when $(\frac{T_0}{T_2})'\neq
0$,\quad $(\frac{T_0}{T_3})'\neq 0$ would give the following four
possibilities:
\begin{eqnarray*}
(d_1)\quad (\frac{T'_0T_2}{T_0T'_2})'\neq 0,\quad
(\frac{T'_0T_3}{T_0T'_3})'\neq 0,\nonumber\\
(d_2)\quad (\frac{T'_0T_2}{T_0T'_2})'\neq 0,\quad
(\frac{T'_0T_3}{T_0T'_3})'=0,\nonumber\\
(d_3)\quad (\frac{T'_0T_2}{T_0T'_2})'=0,\quad
(\frac{T'_0T_3}{T_0T'_3})'\neq 0,\nonumber\\
(d_4)\quad (\frac{T'_0T_2}{T_0T'_2})'=0,\quad
(\frac{T'_0T_3}{T_0T'_3})'=0.
\end{eqnarray*}
In the first possibility 3$a_2b_2c_4(d_1)$, we have either
\begin{eqnarray*}
(e_1)\quad \frac{T_2}{T_3}=constant,\quad (e_2)\quad
\frac{T_2}{T_3}\neq constant.
\end{eqnarray*}
The case 3$a_2b_2c_4d_1(e_1)$ gives one proper MC. The proper MC
is
\begin{equation}
\xi_{(4)}=z\partial_\theta-\frac{T_2}{T_3}\theta\partial_z.
\end{equation}
For the case 3$a_2b_2c_4d_1(e_2)$, we get three MCs which are KVs.
The possibilities 3$a_2b_2c_4(d_2)$ and 3$a_2b_2c_4(d_3)$ are
similar to the case 3$a_2b_2c_4(d_1)$.

In the last possibility 3$a_2b_2c_4(d_4)$, we obtain five MCs in
which two are proper, i.e.,
\begin{eqnarray}
\xi_{(4)}&=&z\partial_\theta-\frac{T_2}{T_3}\theta\partial_z,\nonumber\\
\xi_{(5)}&=&t\partial_t-\frac{2T_0}{T'_0}\partial_r
+\frac{1}{\beta_2}\theta\partial_\theta+\frac{1}{\beta_2}\partial_z,
\end{eqnarray}
where $\frac{T'_0T_2}{T_0T'_2}=\beta_2$, a constant. We also take
$\frac{T_2}{T_3}$ to be constant.

The subcases 3$(a_3)$ and 3$(a_4)$ can be proceeded as the subcase
3$(a_1)$ and would give similar results.

\section{Examples of Finite Dimensional Matter\\
Collineations}

We see from sections 3 and 4 that when we find MCs for the
cylindrically symmetric static spacetimes, we obtain different
constraints on the energy-momentum tensor. If we solve these
constraints, we can have exact solutions of EFEs or class of
solutions can be obtained. In this section, we would attempt to
solve some of these constraints to get explicit form of the
metrics. We are not providing the details rather than we would
provide list of solutions and their properties satisfying the
constraints.\\
\par\noindent
1. When we solve the constraints of the case 2$a_1b_2(c_2)$, we
obtain the following metric
\begin{equation}
ds^2=\cosh^2crdt^2-dr^2-(\cosh cr)^{-1}d\theta^2-(\cosh
cr)^{-1}dz^2,
\end{equation}
where $c$ is an arbitrary constant. This metric admits 4 MCs and
also 4 isometries. This implies that there is no proper MC in this
example but it has 7 RCs. This metric has an anisotropic fluid
with energy-density positive for $0\leq
r<\frac{1}{c}\tanh^{-1}\frac{2}{\sqrt{7}}$ and negative for
$r\geq\frac{1}{c}\tanh^{-1}\frac{2}{\sqrt{7}}$.\\
\par\noindent
2. If we solve the constraints of the case 2$a_2b_3c_1d_3(e_1)$,
the spacetime would be
\begin{equation}
ds^2=(r/r_0)^{2a}dt^2-dr^2-(r/r_0)^{2b}d\theta^2-
(r/r_0)^{2c}dz^2,
\end{equation}
where $a,~b,~c$ and $r_0$ are arbitrary constants such that
$a,b,c\neq 0,1$. The components of the energy-momentum tensor for
this metric are given in Appendix B. This metric has 4 MCs with 3
KVs giving 1 proper MC.\\
\par\noindent
3. If we choose the values of $a,~b,~c$ such that $a=1,~b=c\neq 1$
in Eq.(70), we get a metric which admits 7 MCs and 4 KVs
satisfying the constraints 1$a_4b_1c_2d_1e_2(f_1)$. This gives
three proper MCs. It is obvious from the energy-momentum tensor
given in Eq.(B2) that the energy density will be positive for
$0<b<2/3$.\\
\par\noindent
4. When we take $a=b=c\neq 0,1$ in Eq.(70), we obtain a metric
which satisfies the constraints given in 2$a_1b_1c_2(d_1)$. This
spacetime admits 10 MCs with 6 KVs and hence we have 4 proper MCs
in this example. It can be seen from the energy-momentum tensor
that it becomes singular at $r=0$.\\
\par\noindent
5. If we take $b=c\neq 0,1$ in Eq.(70), we get a metric satisfying
the constraints of the case 2$a_1b_1c_2d_2(e_1)$ admitting 5 MCs
but 3 KVs. This is another example admitting proper MCs. We must
take $0<b<2/3$ to make the energy-density positive. It is to be
noted that the resulting metric would represent a perfect fluid
for $b=a(a-1)/(a+1)$ and non-null electromagnetic field when
$b=a+1$.\\
\par\noindent
6. Taking $\nu=\lambda=\mu$ in Eq.(4), it satisfies the
constraints of the case 2$a_1b_2(c_1)$. This metric admits 6 MCs
and also 6 KVs.\\
\par\noindent
7. When we choose $a=b\neq c$ such that $a,c\neq 0,1$ in Eq.(70),
we obtain a metric satisfying the constraints of the case
2$a_2b_3c_1(d_1)$. This metric admits 5 MCs with 4 isometries
hence giving 1 proper MC.\\
\par\noindent
8. If we take either $\nu=\lambda$ or $\nu=\mu$ in Eq.(4), we have
the solution of the constraints given by 2$a_2b_3c_2(d_1)$.
This metric admits 4 MCs and 4 isometries.\\
\par\noindent
9. Now solving the constraints of the case 2$a_2b_3c_2(d_2)$, we
obtain the following solution
\begin{equation}
ds^2=(\cosh cr)^{-1}dt^2-dr^2-\cosh^2crd\theta^2-(\cosh
cr)^{-1}dz^2.
\end{equation}
It has 4 MCs and also 4 KVs but 7 RCs. This spacetime represents
anisotropic tachyonic fluid.\\
\par\noindent
10. If we take $a=-\frac{1}{4},~b=c=\frac{1}{2}$ in Eq.(70), we
get $T_1=0$ which gives the degenerate case and hence satisfies
the constraints of the case 3$a_2b_2c_4(d_4)$. This metric admits
5MCs and 4 KVs. Thus we have one proper MC in this degenerate
case.

\section{Conclusion}

We know from the classification of cylindrically symmetric static
spacetimes according to their isometries that we either get three,
four, five, six, seven or ten isometries. The ten and seven KVs
are admitted by the well known anti-de Sitter and anti-Einstein
universes respectively. The six isometries are admitted by the
Bertotti-Robinson metric with the isometry group
$SO(1,1)\times)\Re^3\otimes SO(3)$. There is one class of metrics
depending on one arbitrary function with six isometries given by
\begin{equation}
ds^2=e^\nu(dt^2-dr^2-d\theta^2-dz^2).
\end{equation}
There are three cases of five dimensional isometry groups. There
are also three classes of metrics depending upon two arbitrary
functions having four dimensional isometry groups given by
\begin{equation}
ds^2=e^\nu dt^2-dr^2-e^\mu(d\theta^2-dz^2),
\end{equation}
\begin{equation}
ds^2=e^\nu(dt^2-dz^2)-dr^2-e^\lambda d\theta^2,
\end{equation}
\begin{equation}
ds^2=e^\nu(dt^2-d\theta^2)-dr^2-e^\mu dz^2.
\end{equation}
All other metrics admit minimal isometry group $G_3$.

This paper presents a complete classification of cylindrically
symmetric static spacetimes according to their MCs. We have solved
MC equations for both non-degenerate and degenerate cases. The
explicit forms of MCs are given in each case. We have also written
the corresponding constraints on the energy-momentum tensor in
each case. Finally, we have attempted to solve some of these
constraints to find the exact solution of EFEs.

When the energy-momentum tensor is non-degenerate (section 3), we
obtain either three, four, five, six, seven or ten independent
MCs. Out of these MCs, we obtain three isometries and the rest are
the the non-trivial (proper) MCs. In the degenerate case (section
4), most of the possibilities lead to infinite dimensional MCs.
However, there are some worth mentioning cases where we obtain
finite dimensional MCs even with degenerate energy-momentum
tensor. In these case, we get either three, four, five or ten
independent MCs in which three are the usual KVs and the rest are
the proper MCs.

The summary of the results can be given below in the form of
tables.
\vspace{0.2cm}

{\bf {\small Table 1.}} {\small MCs of Case (1) for the
Non-degenerate Energy-Momentum Tensor}.

\vspace{0.1cm}

\begin{center}
\begin{tabular}{|l|l|l|}
\hline {\bf Cases} & {\bf MCs} & {\bf Constraints}
\\ \hline 1$(a_1)$ & $10$ & $T'_0=0,~T'_2=0=T'_3$
\\ \hline 1$a_2b_1(c_1)$ & $6$ &$
\begin{array}{c}
T'_0=0,~T'_2=0,~T'_3\neq 0,~
[\frac{T_3}{\sqrt{T_1}}(\frac{T'_3}{2T_3\sqrt{T_1}})']'=0,\\
\frac{T_3}{\sqrt{T_1}}(\frac{T'_3}{2T_3\sqrt{T_1}})'=\alpha_1>0
\end{array}
$\\ \hline 1$a_2b_1(c_2)$ & $6$ &$
\begin{array}{c}
T'_0=0,~T'_2=0,~T'_3\neq 0,~
[\frac{T_3}{\sqrt{T_1}}(\frac{T'_3}{2T_3\sqrt{T_1}})']'=0,\\
\alpha_1=0
\end{array}
$\\ \hline 1$a_2b_1(c_3)$ & $6$&$
\begin{array}{c}
T'_0=0,~T'_2=0,~T'_3\neq 0,~
[\frac{T_3}{\sqrt{T_1}}(\frac{T'_3}{2T_3\sqrt{T_1}})']'=0,\\
\alpha_1<0
\end{array}
$\\ \hline 1$a_2(b_2)$ & $4$ &$
\begin{array}{c}
T'_0=0,~T'_2=0,~T'_3\neq 0,~
[\frac{T_3}{\sqrt{T_1}}(\frac{T'_3}{2T_3\sqrt{T_1}})']'\neq 0
\end{array}
$\\ \hline
\end{tabular}
\end{center}
\begin{center}
\begin{tabular}{|l|l|l|}
\hline {\bf Cases} & {\bf MCs} & {\bf Constraints}
\\ \hline 1$(a_3)$ & $1(a_2)$ & $T'_0=0,~T'_2\neq 0,~T'_3=0$
\\ \hline 1$a_4b_1c_1(d_1)$ & $6$ & $
\begin{array}{c}
T'_0=0,~T'_2\neq 0,~T'_3\neq 0,~
[\frac{T_2}{\sqrt{T_1}}(\frac{T'_2}{2T_2\sqrt{T_1}})']'=0,\\
\frac{T_2}{\sqrt{T_1}}(\frac{T'_2}{2T_2\sqrt{T_1}})'=\alpha_3>0,~
[\frac{T_3}{\sqrt{T_1}}(\frac{T'_3}{2T_3\sqrt{T_1}})']'=0
\end{array}
$\\ \hline 1$a_4b_1c_1d_2(e_1)$ & $4$ & $
\begin{array}{c}
T'_0=0,~T'_2\neq 0,~T'_3\neq 0,~
[\frac{T_2}{\sqrt{T_1}}(\frac{T'_2}{2T_2\sqrt{T_1}})']'=0,\\
\alpha_3>0,~
[\frac{T_3}{\sqrt{T_1}}(\frac{T'_3}{2T_3\sqrt{T_1}})']'\neq 0,~
\frac{T_2}{T_3}=constant\neq 0
\end{array}
$\\ \hline 1$a_4b_1c_1d_2(e_2)$ & $3$ & $
\begin{array}{c}
T'_0=0,~T'_2\neq 0,~T'_3\neq 0,~
[\frac{T_2}{\sqrt{T_1}}(\frac{T'_2}{2T_2\sqrt{T_1}})']'=0,\\
\alpha_3>0,~
[\frac{T_3}{\sqrt{T_1}}(\frac{T'_3}{2T_3\sqrt{T_1}})']'\neq 0,~
\frac{T_2}{T_3}\neq constant
\end{array}
$\\ \hline 1$a_4b_1c_2d_1(e_1)$ & $6$ & $
\begin{array}{c}
T'_0=0,~T'_2\neq 0,~T'_3\neq 0,~
[\frac{T_2}{\sqrt{T_1}}(\frac{T'_2}{2T_2\sqrt{T_1}})']'=0,\\
\alpha_3=0,~
[\frac{T_3}{\sqrt{T_1}}(\frac{T'_3}{2T_3\sqrt{T_1}})']'=0,~
\alpha_1>0
\end{array}
$\\ \hline 1$a_4b_1c_2d_1e_2(f_1)$ & $7$ & $
\begin{array}{c}
T'_0=0,~T'_2\neq 0,~T'_3\neq 0,~
[\frac{T_2}{\sqrt{T_1}}(\frac{T'_2}{2T_2\sqrt{T_1}})']'=0,\\
\alpha_3=0,~
[\frac{T_3}{\sqrt{T_1}}(\frac{T'_3}{2T_3\sqrt{T_1}})']'=0,~
\alpha_1=0,\\
\frac{T_3}{T_2}=constant
\end{array}
$\\ \hline 1$a_4b_1c_2d_1e_2(f_2)$ & $4$ & $
\begin{array}{c}
T'_0=0,~T'_2\neq 0,~T'_3\neq 0,~
[\frac{T_2}{\sqrt{T_1}}(\frac{T'_2}{2T_2\sqrt{T_1}})']'=0,\\
\alpha_3=0,~
[\frac{T_3}{\sqrt{T_1}}(\frac{T'_3}{2T_3\sqrt{T_1}})']'=0,
\alpha_1=0,\\
\frac{T_3}{T_2}\neq constant
\end{array}
$\\ \hline 1$a_4b_1c_2d_1(e_3)$ & $6$ & $
\begin{array}{c}
T'_0=0,~T'_2\neq 0,~T'_3\neq 0,~
[\frac{T_2}{\sqrt{T_1}}(\frac{T'_2}{2T_2\sqrt{T_1}})']'=0,\\
\alpha_3=0,~
[\frac{T_3}{\sqrt{T_1}}(\frac{T'_3}{2T_3\sqrt{T_1}})']'=0,~
\alpha_1<0
\end{array}
$\\ \hline 1$a_4b_1c_2(d_2)$ & $4$ & $
\begin{array}{c}
T'_0=0,~T'_2\neq 0,~T'_3\neq 0,~
[\frac{T_2}{\sqrt{T_1}}(\frac{T'_2}{2T_2\sqrt{T_1}})']'=0,\\
\alpha_3=0,~
[\frac{T_3}{\sqrt{T_1}}(\frac{T'_3}{2T_3\sqrt{T_1}})']'\neq 0
\end{array}
$\\ \hline 1$a_4b_1(c_3)$ & $6$ & $
\begin{array}{c}
T'_0=0,~T'_2\neq 0,~T'_3\neq 0,~
[\frac{T_2}{\sqrt{T_1}}(\frac{T'_2}{2T_2\sqrt{T_1}})']'=0,\\
\alpha_3<0
\end{array}
$\\ \hline 1$a_4b_2(c_1)$ & $4$ & $
\begin{array}{c}
T'_0=0,~T'_2\neq 0,~T'_3\neq 0,~
[\frac{T_2}{\sqrt{T_1}}(\frac{T'_2}{2T_2\sqrt{T_1}})']'\neq 0,\\
\frac{T_2}{T_3}=constant
\end{array}
$\\ \hline 1$a_4b_2(c_2)$ & $3$ & $
\begin{array}{c}
T'_0=0,~T'_2\neq 0,~T'_3\neq 0,~
[\frac{T_2}{\sqrt{T_1}}(\frac{T'_2}{2T_2\sqrt{T_1}})']'\neq 0,\\
\frac{T_2}{T_3}\neq constant
\end{array}
$\\ \hline
\end{tabular}
\end{center}
\newpage

{\bf {\small Table 2}. }{\small MCs of Case (2) for the
Non-degenerate Energy-Momentum Tensor}.

\vspace{0.1cm}

\begin{center}
\begin{tabular}{|l|l|l|}
\hline {\bf Cases} & {\bf MCs} & {\bf Constraints}
\\ \hline 2$a_1b_1c_1(d_1)$ & $10$ & $
\begin{array}{c}
T'_0\neq
0,~(\frac{T_2}{T_3})'=0,~(\frac{T'_2}{2T_2\sqrt{T_1}})'=0,\\
\frac{T'_2}{2T_2\sqrt{T_1}}=\beta_1=0,~
(\frac{(\sqrt{T_0})'}{\sqrt{T_1}})'=0
\end{array}
$\\ \hline 2$a_1b_1c_1d_2e_1(f_1)$ & $6$ &$
\begin{array}{c}
T'_0\neq
0,~(\frac{T_2}{T_3})'=0,~(\frac{T'_2}{2T_2\sqrt{T_1}})'=0,~
\beta_1=0,\\(\frac{(\sqrt{T_0})'} {\sqrt{T_1}})'\neq 0,~
\frac{T_0}{2\sqrt{T_1}}(\frac{T'_0}{T_0\sqrt{T_1}})'=\beta_2
=constant,\\\beta_2=0
\end{array}
$\\ \hline 2$a_1b_1c_1d_2e_1(f_2)$ & $6$ &$
\begin{array}{c}
T'_0\neq
0,~(\frac{T_2}{T_3})'=0,~(\frac{T'_2}{2T_2\sqrt{T_1}})'=0,~
\beta_1=0,\\(\frac{(\sqrt{T_0})'} {\sqrt{T_1}})'\neq 0,~
\beta_2=constant,~\beta_2\neq 0
\end{array}
$\\ \hline 2$a_1b_1c_1d_2(e_2)$ & $4$ &$
\begin{array}{c}
T'_0\neq
0,~(\frac{T_2}{T_3})'=0,~(\frac{T'_2}{2T_2\sqrt{T_1}})'=0,\\
\beta_1=0,~(\frac{(\sqrt{T_0})'} {\sqrt{T_1}})'\neq 0,~
\beta_2\neq constant
\end{array}
$\\ \hline 2$a_1b_1c_2(d_1)$ & $10$ &$
\begin{array}{c}
T'_0\neq
0,~(\frac{T_2}{T_3})'=0,~(\frac{T'_2}{2T_2\sqrt{T_1}})'=0,~
\beta_1\neq 0,
\\(\frac{T_2}{T_0})'=0
\end{array}
$\\ \hline 2$a_1b_1c_2d_2(e_1)$ & $5$ & $
\begin{array}{c} T'_0\neq
0,~(\frac{T_2}{T_3})'=0,~(\frac{T'_2}{2T_2\sqrt{T_1}})'=0,\\
\beta_1\neq 0,~(\frac{T_2}{T_0})'\neq
0,~(\frac{T'_0}{T_0\sqrt{T_1}})'=0
\end{array}$
\\ \hline 2$a_1b_1c_2d_2(e_2)$ & $4$ &
$\begin{array}{c}
T'_0\neq 0,~(\frac{T_2}{T_3})'=0,~(\frac{T'_2}{2T_2\sqrt{T_1}})'=0,\\
\beta_1\neq 0,~(\frac{T_2}{T_0})'\neq
0,~(\frac{T'_0}{T_0\sqrt{T_1}})'\neq 0
\end{array}$
\\ \hline 2$a_1b_2(c_1)$ & $6$ &
$\begin{array}{c}
T'_0\neq 0,~(\frac{T_2}{T_3})'=0,~(\frac{T'_2}
{2T_2\sqrt{T_1}})'\neq 0,\\
(\sqrt{\frac{T_0}{T_2}})'=0
\end{array}$
\\ \hline 2$a_1b_2(c_2)$ & $4$ &
$\begin{array}{c}
T'_0\neq 0,~(\frac{T_2}{T_3})'=0,~(\frac{T'_2}
{2T_2\sqrt{T_1}})'\neq 0,\\
(\sqrt{\frac{T_0}{T_2}})'\neq 0
\end{array}
$\\ \hline 2$a_2b_1c_1(d_1)$ & $7$ &$
\begin{array}{c}
T'_0\neq 0,~(\frac{T_2}{T_3})'\neq 0,~T'_2=0,~T'_3\neq
0,\\
\frac{T_0}{\sqrt{T_1}}(\frac{T'_0}{2T_0\sqrt{T_1}})'=\gamma_1=0,\\
\frac{T_3}{\sqrt{T_1}}(\frac{T'_3}{2T_3\sqrt{T_1}})'=\gamma_2=0,~
({\frac{T_0}{T_3}})'=0
\end{array}
$\\ \hline 2$a_2b_1c_1(d_2)$ & $4$ &$
\begin{array}{c}
T'_0\neq 0,~(\frac{T_2}{T_3})'\neq 0,~T'_2=0,~T'_3\neq 0,\\
\gamma_1=0,~\gamma_2=0,~ ({\frac{T_0}{T_3}})'\neq 0
\end{array}
$\\ \hline 2$a_2b_1c_2(d_1)$ & $3$ & $
\begin{array}{c}
T'_0\neq 0,~(\frac{T_2}{T_3})'\neq 0,~T'_2=0,~T'_3\neq 0,\\
\gamma_1\neq 0,~\gamma_2=0,~\gamma_1>0
\end{array}
$\\ \hline
\end{tabular}
\end{center}
\begin{center}
\begin{tabular}{|l|l|l|}
\hline {\bf Cases} & {\bf MCs} & {\bf Constraints}
\\ \hline 2$a_2b_1c_2d_2(e_1)$ & $4$ & $
\begin{array}{c}
T'_0\neq 0,~(\frac{T_2}{T_3})'\neq 0,~T'_2=0,~T'_3\neq 0,\\
\gamma_1\neq 0,~\gamma_2=0,~\gamma_1<0,~({\frac{T_0}{T_3}})'=0
\end{array}
$\\ \hline 2$a_2b_1c_2d_2(e_2)$ & $3$ & $
\begin{array}{c}
T'_0\neq 0,~(\frac{T_2}{T_3})'\neq 0,~T'_2=0,~T'_3\neq 0,\\
\gamma_1\neq 0,~\gamma_2=0,~\gamma_1<0,~({\frac{T_0}{T_3}})'\neq 0
\end{array}
$\\ \hline 2$a_2b_1(c_3)$ & 2$a_2b_1(c_2)$ & $
\begin{array}{c}
T'_0\neq 0,~(\frac{T_2}{T_3})'\neq 0,~T'_2=0,~T'_3\neq 0,\\
\gamma_1=0,~\gamma_2\neq 0
\end{array}
$\\ \hline 2$a_2b_1c_4d_1(e_1)$ & $3$ & $
\begin{array}{c}
T'_0\neq 0,~(\frac{T_2}{T_3})'\neq 0,~T'_2=0,~T'_3\neq 0,\\
\gamma_1\neq 0,~\gamma_2\neq 0,~\gamma_1>0,~\gamma_2>0,\\
\gamma_2T_0\int{\frac{\sqrt{T_1}}{T_0}}dr+\frac{T'_3}{2\sqrt{T_1}}=0
\end{array}
$\\ \hline 2$a_2b_1c_4d_1e_2(f_1)$ & $4$ & $
\begin{array}{c}
T'_0\neq 0,~(\frac{T_2}{T_3})'\neq 0,~T'_2=0,~T'_3\neq 0,\\
\gamma_1\neq 0,~\gamma_2\neq 0,~\gamma_1>0,~\gamma_2>0,\\
\gamma_2T_0\int{\frac{\sqrt{T_1}}{T_0}}dr+\frac{T'_3}{2\sqrt{T_1}}\neq
0,~(\frac{T_0}{T_3})'=0
\end{array}
$\\ \hline 2$a_2b_1c_4d_1e_2(f_2)$ & $3$ & $
\begin{array}{c}
T'_0\neq 0,~(\frac{T_2}{T_3})'\neq 0,~T'_2=0,~T'_3\neq 0,\\
\gamma_1\neq 0,~\gamma_2\neq 0,~\gamma_1>0,~\gamma_2>0,\\
\gamma_2T_0\int{\frac{\sqrt{T_1}}{T_0}}dr+\frac{T'_3}{2\sqrt{T_1}}\neq
0,~(\frac{T_0}{T_3})'\neq 0
\end{array}
$\\ \hline 2$a_2b_1c_4(d_2)$ & 2$a_2b_1c_4(d_1)$ &$
\begin{array}{c}
T'_0\neq 0,~(\frac{T_2}{T_3})'\neq 0,~T'_2=0,~T'_3\neq 0,\\
\gamma_1\neq 0,~\gamma_2\neq 0,~\gamma_1>0,~\gamma_2<0
\end{array}
$\\ \hline 2$a_2b_1c_4(d_3)$ & 2$a_2b_1c_4(d_1)$ &$
\begin{array}{c}
T'_0\neq 0,~(\frac{T_2}{T_3})'\neq 0,~T'_2=0,~T'_3\neq 0,\\
\gamma_1\neq 0,~\gamma_2\neq 0,~\gamma_1<0,~\gamma_2>0
\end{array}
$\\ \hline 2$a_2b_1c_4(d_4)$ & 2$a_2b_1c_4(d_1)$ &$
\begin{array}{c}
T'_0\neq 0,~(\frac{T_2}{T_3})'\neq 0,~T'_2=0,~T'_3\neq 0,\\
\gamma_1\neq 0,~\gamma_2\neq 0,~\gamma_1<0,~\gamma_2<0
\end{array}
$\\ \hline 2$a_2(b_2)$ & 2$a_2(b_1)$ & $
\begin{array}{c}
T'_0\neq 0,~(\frac{T_2}{T_3})'\neq 0,~T'_2\neq 0,~T'_3=0
\end{array}
$\\ \hline 2$a_2b_3c_1(d_1)$ & $5$ & $
\begin{array}{c}
T'_0\neq 0,~(\frac{T_2}{T_3})'\neq 0,~T'_2\neq 0,~T'_3\neq 0,\\
(\frac{T'_2}{2T_2\sqrt{T_1}})'=0,~(\frac{T'_3}{2T_3\sqrt{T_1}})'=0,~
(\frac{T_2}{T_0})'=0,\\
(\frac{T_3}{T_0})'\neq 0
\end{array}
$\\ \hline
\end{tabular}
\end{center}
\begin{center}
\begin{tabular}{|l|l|l|}
\hline {\bf Cases} & {\bf MCs} & {\bf Constraints}
\\ \hline 2$a_2b_3c_1(d_2)$ & $5$ & $
\begin{array}{c}
T'_0\neq 0,~(\frac{T_2}{T_3})'\neq 0,~T'_2\neq 0,~T'_3\neq 0,\\
(\frac{T'_2}{2T_2\sqrt{T_1}})'=0,~(\frac{T'_3}{2T_3\sqrt{T_1}})'=0,~
(\frac{T_2}{T_0})'\neq 0,\\
(\frac{T_3}{T_0})'=0
\end{array}
$\\ \hline 2$a_2b_3c_1d_3(e_1)$ & $4$ & $
\begin{array}{c}
T'_0\neq 0,~(\frac{T_2}{T_3})'\neq 0,~T'_2\neq 0,~T'_3\neq 0,\\
(\frac{T'_2}{2T_2\sqrt{T_1}})'=0,~(\frac{T'_3}{2T_3\sqrt{T_1}})'=0,~
(\frac{T_2}{T_0})'\neq 0,\\
(\frac{T_3}{T_0})'\neq 0,~(\frac{T'_0}{2T_0\sqrt{T_1}})'=0
\end{array}
$\\ \hline 2$a_2b_3c_1d_3(e_2)$ & $3$ & $
\begin{array}{c}
T'_0\neq 0,~(\frac{T_2}{T_3})'\neq 0,~T'_2\neq 0,~T'_3\neq 0,\\
(\frac{T'_2}{2T_2\sqrt{T_1}})'=0,~(\frac{T'_3}{2T_3\sqrt{T_1}})'=0,~
(\frac{T_2}{T_0})'\neq 0,\\
(\frac{T_3}{T_0})'\neq 0,~(\frac{T'_0}{2T_0\sqrt{T_1}})'\neq 0
\end{array}
$\\ \hline 2$a_2b_3c_2(d_1)$ & $4$ & $
\begin{array}{c}
T'_0\neq 0,~(\frac{T_2}{T_3})'\neq 0,~T'_2\neq 0,~T'_3\neq 0,\\
(\frac{T'_2}{2T_2\sqrt{T_1}})'\neq 0,~
(\frac{T'_3}{2T_3\sqrt{T_1}})'\neq 0,~
(\frac{T_2}{T_0})'=0,\\
(\frac{T_3}{T_0})'\neq 0
\end{array}
$\\ \hline 2$a_2b_3c_2(d_2)$ & $4$ & $
\begin{array}{c}
T'_0\neq 0,~(\frac{T_2}{T_3})'\neq 0,~T'_2\neq 0,~T'_3\neq 0,\\
(\frac{T'_2}{2T_2\sqrt{T_1}})'\neq 0,~
(\frac{T'_3}{2T_3\sqrt{T_1}})'\neq 0,~
(\frac{T_2}{T_0})'\neq 0,\\
(\frac{T_3}{T_0})'=0
\end{array}
$\\ \hline 2$a_2b_3c_2(d_3)$ & $3$ & $
\begin{array}{c}
T'_0\neq 0,~(\frac{T_2}{T_3})'\neq 0,~T'_2\neq 0,~T'_3\neq 0,\\
(\frac{T'_2}{2T_2\sqrt{T_1}})'\neq 0,~
(\frac{T'_3}{2T_3\sqrt{T_1}})'\neq 0,~
(\frac{T_2}{T_0})'\neq 0,\\
(\frac{T_3}{T_0})'\neq 0
\end{array}
$\\ \hline
\end{tabular}
\end{center}
\newpage

{\bf {\small Table 3}. }{\small MCs of Case (1) for the Degenerate
Energy-Momentum Tensor}.

\vspace{0.1cm}

\begin{center}
\begin{tabular}{|l|l|l|}
\hline {\bf Cases} & {\bf MCs} & {\bf Constraints}
\\ \hline 1$(a_1)$ & Infinite No. of MCs & $T_0=0,~T_1=0,~
T_2=0,~T_3\neq 0$
\\ \hline $1(a_2)$ &Infinite No. of MCs&
$T_0=0,~T_1=0,~ T_2\neq 0,~T_3=0$
\\ \hline $1(a_3)$ & Infinite No. of MCs &
$T_0=0,~T_1\neq 0,~T_2=0,~T_3=0$
\\ \hline $1(a_4)$ & Infinite No. of MCs &
$T_0\neq 0,~T_1=0,~T_2\neq 0,~T_3=0$
\\ \hline
\end{tabular}
\end{center}
\vspace{0.2cm}

{\bf {\small Table 4}. }{\small MCs of Case (2) for the Degenerate
Energy-Momentum Tensor}.

\vspace{0.1cm}

\begin{center}
\begin{tabular}{|l|l|l|}
\hline {\bf Cases} & {\bf MCs} & {\bf Constraints}
\\ \hline 2$a_1(b_1)$ & Infinite No. of MCs &$
\begin{array}{c}
T_0=0,~T_1=0,~T_2\neq 0,~T_3\neq 0,\\
(\frac{T_3}{T_2})'=0
\end{array}
$\\ \hline 2$a_1b_2(c_1)$ & Infinite No. of MCs & $
\begin{array}{c}
T_0=0,~T_1=0,~T_2\neq 0,~T_3\neq 0,\\
(\frac{T_3}{T_2})'\neq 0,~T'_2=0,~T'_3\neq 0
\end{array}
$\\ \hline 2$a_1b_2(c_2)$ &Infinite No. of MCs& $
\begin{array}{c}
T_0=0,~T_1=0,~T_2\neq 0,~T_3\neq 0,\\
(\frac{T_3}{T_2})'\neq 0,~T'_2\neq 0,~T'_3=0
\end{array}
$\\ \hline 2$a_1b_2(c_3)$ &Infinite No. of MCs& $
\begin{array}{c}
T_0=0,~T_1=0,~T_2\neq 0,~T_3\neq 0,\\
(\frac{T_3}{T_2})'\neq 0,~T'_2\neq 0,~T'_3\neq 0
\end{array}
$\\ \hline 2$a_2b_1(c_1)$ &Infinite No. of MCs& $
\begin{array}{c}
T_0=0,~T_1\neq 0,~T_2=0,~T_3\neq 0,\\
(\frac{T_3}{\sqrt{T_1}}(\frac{T'_3}{2T_3\sqrt{T_1}})')'=0,\\
\frac{T_3}{\sqrt{T_1}}(\frac{T'_3}{2T_3\sqrt{T_1}})'=c>0
\end{array}
$\\ \hline 2$a_2b_1(c_2)$ &Infinite No. of MCs& $
\begin{array}{c}
T_0=0,~T_1\neq 0,~T_2=0,~T_3\neq 0,\\
(\frac{T_3}{\sqrt{T_1}}(\frac{T'_3}{2T_3\sqrt{T_1}})')'=0,~c=0
\end{array}
$\\ \hline 2$a_2b_1(c_3)$ &Infinite No. of MCs& $
\begin{array}{c}
T_0=0,~T_1\neq 0,~T_2=0,~T_3\neq 0,\\
(\frac{T_3}{\sqrt{T_1}}(\frac{T'_3}{2T_3\sqrt{T_1}})')'=0,~c<0
\end{array}
$\\ \hline 2$a_2(b_2)$ &Infinite No. of MCs& $
\begin{array}{c}
T_0=0,~T_1\neq 0,~T_2=0,~T_3\neq 0,\\
(\frac{T_3}{\sqrt{T_1}}(\frac{T'_3}{2T_3\sqrt{T_1}})')'\neq 0
\end{array}
$\\ \hline 2$(a_3)$ &Infinite No. of MCs& $
\begin{array}{c}
T_0=0,~T_1\neq 0,~T_2\neq 0,~T_3=0
\end{array}
$\\ \hline 2$(a_4)$ &Infinite No. of MCs& $
\begin{array}{c}
T_0\neq 0,~T_1= 0,~T_2=0,~T_3\neq 0
\end{array}
$\\ \hline 2$(a_5)$ &Infinite No. of MCs& $
\begin{array}{c}
T_0\neq 0,~T_1= 0,~T_2\neq 0,~T_3=0
\end{array}
$\\ \hline 2$(a_6)$ &Infinite No. of MCs& $
\begin{array}{c}
T_0\neq 0,~T_1\neq 0,~T_2=0,~T_3=0
\end{array}
$\\ \hline
\end{tabular}
\end{center}
\newpage

{\bf {\small Table 5}. }{\small MCs of Case (3) for the Degenerate
Energy-Momentum Tensor}.

\vspace{0.1cm}

\begin{center}
\begin{tabular}{|l|l|l|}
\hline {\bf Cases} & {\bf MCs} & {\bf Constraints}
\\ \hline 3$a_1b_1c_1d_1(e_1)$ &Infinite No. of MCs& $
\begin{array}{c}
T_0=0,~T_1\neq 0,~T_2\neq 0,~T_3\neq 0,\\
(\frac{T_2}{\sqrt{T_1}}(\frac{T'_2}{2T_2\sqrt{T_1}})')'=0,\\
\frac{T_2}{\sqrt{T_1}}(\frac{T'_2}{2T_2\sqrt{T_1}})'=\alpha_1>0,\\
\frac{T_3}{\sqrt{T_1}}(\frac{T'_3}{2T_3\sqrt{T_1}})'=\alpha_2>0,\\
\sqrt{\frac{\alpha_1}{\alpha_2}}+\sqrt{\frac{\alpha_2}{\alpha_1}}=0
\end{array}
$\\ \hline 3$a_1b_1c_1d_1(e_2)$ &Infinite No. of MCs& $
\begin{array}{c}
T_0=0,~T_1\neq 0,~T_2\neq 0,~T_3\neq 0,\\
(\frac{T_2}{\sqrt{T_1}}(\frac{T'_2}{2T_2\sqrt{T_1}})')'=0,\\
\alpha_1>0,~\alpha_2>0,~
\sqrt{\frac{\alpha_1}{\alpha_2}}+\sqrt{\frac{\alpha_2}{\alpha_1}}\neq
0
\end{array}
$\\ \hline 3$a_1b_1c_1(d_2)$ &Infinite No. of MCs& $
\begin{array}{c}
T_0=0,~T_1\neq 0,~T_2\neq 0,~T_3\neq 0,\\
(\frac{T_2}{\sqrt{T_1}}(\frac{T'_2}{2T_2\sqrt{T_1}})')'=0,\\
\alpha_1>0,~\alpha_2=0
\end{array}
$\\ \hline 3$a_1b_1c_1(d_3)$ &Infinite No. of MCs&$
\begin{array}{c}
T_0=0,~T_1\neq 0,~T_2\neq 0,~T_3\neq 0,\\
(\frac{T_2}{\sqrt{T_1}}(\frac{T'_2}{2T_2\sqrt{T_1}})')'=0,\\
\alpha_1>0,~ \alpha_2<0
\end{array}
$\\ \hline 3$a_1b_1c_2d_1(e_1)$ &Infinite No. of MCs&$
\begin{array}{c}
T_0=0,~T_1\neq 0,~T_2\neq 0,~T_3\neq 0,\\
(\frac{T_2}{\sqrt{T_1}}(\frac{T'_2}{2T_2\sqrt{T_1}})')'=0,~
\alpha_1=0,\\
-\frac{T'_2}{2T_2\sqrt{T_1}}=\alpha_4=0,~\alpha_2>0
\end{array}
$\\ \hline
\end{tabular}
\end{center}
\begin{center}
\begin{tabular}{|l|l|l|}
\hline {\bf Cases} & {\bf MCs} & {\bf Constraints}
\\ \hline 3$a_1b_1c_2d_1e_2(f_1)$ &Infinite No. of MCs&$
\begin{array}{c}
T_0=0,~T_1\neq 0,~T_2\neq 0,~T_3\neq 0,\\
(\frac{T_2}{\sqrt{T_1}}(\frac{T'_2}{2T_2\sqrt{T_1}})')'=0,~
\alpha_1=0,\\
\alpha_4=0,~\alpha_2=0,\\
-\frac{T'_3}{2T_3\sqrt{T_1}}=\alpha_5=0
\end{array}
$\\ \hline 3$a_1b_1c_2d_1e_2(f_2)$ &Infinite No. of MCs& $
\begin{array}{c}
T_0=0,~T_1\neq 0,~T_2\neq 0,~T_3\neq 0,\\
(\frac{T_2}{\sqrt{T_1}}(\frac{T'_2}{2T_2\sqrt{T_1}})')'=0,~
\alpha_1=0,\\
\alpha_4=0,~\alpha_2=0,~\alpha_5\neq 0
\end{array}
$\\ \hline 3$a_1b_1c_2d_1(e_3)$ &Infinite No. of MCs&$
\begin{array}{c}
T_0=0,~T_1\neq 0,~T_2\neq 0,~T_3\neq 0,\\
(\frac{T_2}{\sqrt{T_1}}(\frac{T'_2}{2T_2\sqrt{T_1}})')'=0,~
\alpha_1=0,\\
\alpha_4=0,~\alpha_2<0
\end{array}
$\\ \hline 3$a_1b_1c_2d_2(e_1)$ &Infinite No. of MCs&$
\begin{array}{c}
T_0=0,~T_1\neq 0,~T_2\neq 0,~T_3\neq 0,\\
(\frac{T_2}{\sqrt{T_1}}(\frac{T'_2}{2T_2\sqrt{T_1}})')'=0,~
\alpha_1=0,\\
\alpha_4\neq 0,~\alpha_2>0
\end{array}
$\\ \hline 3$a_1b_1c_2d_2(e_2)$ &Infinite No. of MCs& $
\begin{array}{c}
T_0=0,~T_1\neq 0,~T_2\neq 0,~T_3\neq 0,\\
(\frac{T_2}{\sqrt{T_1}}(\frac{T'_2}{2T_2\sqrt{T_1}})')'=0,~
\alpha_1=0,\\
\alpha_4\neq 0,~\alpha_2=0
\end{array}
$\\ \hline 3$a_1b_1c_2d_2(e_3)$ &Infinite No. of MCs&$
\begin{array}{c}
T_0=0,~T_1\neq 0,~T_2\neq 0,~T_3\neq 0,\\
(\frac{T_2}{\sqrt{T_1}}(\frac{T'_2}{2T_2\sqrt{T_1}})')'=0,~
\alpha_1=0,\\
\alpha_4\neq 0,~\alpha_2<0
\end{array}
$\\ \hline 3$a_1b_1(c_3)$ &Infinite No. of MCs&$
\begin{array}{c}
T_0=0,~T_1\neq 0,~T_2\neq 0,~T_3\neq 0,\\
(\frac{T_2}{\sqrt{T_1}}(\frac{T'_2}{2T_2\sqrt{T_1}})')'=0,~
\alpha_1<0,
\end{array}
$\\ \hline
\end{tabular}
\end{center}
\begin{center}
\begin{tabular}{|l|l|l|}
\hline {\bf Cases} & {\bf MCs} & {\bf Constraints}
\\ \hline 3$a_1(b_2)$ &Infinite No. of MCs& $
\begin{array}{c}
T_0=0,~T_1\neq 0,~T_2\neq 0,~T_3\neq 0,\\
(\frac{T_2}{\sqrt{T_1}}(\frac{T'_2}{2T_2\sqrt{T_1}})')'\neq 0
\end{array}
$\\ \hline 3$a_2b_1(c_1)$ & $4$ & $
\begin{array}{c}
T_0\neq 0,~T_1=0,~T_2\neq 0,~T_3\neq 0,\\
T'_0=0,~(\frac{T'_2T_3}{T_2T'_3})'=0
\end{array}
$\\ \hline 3$a_2b_1c_2(d_1)$ & Infinite No. of MCs & $
\begin{array}{c}
T_0\neq 0,~T_1=0,~T_2\neq 0,~T_3\neq 0,\\
T'_0=0,~(\frac{T'_2T_3}{T_2T'_3})'\neq 0,~T'_2=0=T'_3
\end{array}
$\\ \hline 3$a_2b_1c_2(d_2)$ & $3$ & $
\begin{array}{c}
T_0\neq 0,~T_1=0,~T_2\neq 0,~T_3\neq 0,\\
T'_0=0,~(\frac{T'_2T_3}{T_2T'_3})'\neq 0,~T'_2\neq 0,\\
T'_3\neq 0
\end{array}
$\\ \hline 3$a_2b_2(c_1)$ & $10$ & $
\begin{array}{c}
T_0\neq 0,~T_1=0,~T_2\neq 0,~T_3\neq 0,\\
T'_0\neq 0,~(\frac{T_0}{T_2})'=0,~(\frac{T_0}{T_3})'=0
\end{array}
$\\ \hline 3$a_2b_2c_2(d_1)$ & $5$ & $
\begin{array}{c}
T_0\neq 0,~T_1=0,~T_2\neq 0,~T_3\neq 0,\\
T'_0\neq 0,~(\frac{T_0}{T_2})'=0,~(\frac{T_0}{T_3})'\neq 0,\\
(\frac{T_0}{T_2})'=0
\end{array}
$\\ \hline 3$a_2b_2c_2(d_2)$ & $4$ & $
\begin{array}{c}
T_0\neq 0,~T_1=0,~T_2\neq 0,~T_3\neq 0,\\
T'_0\neq 0,~(\frac{T_0}{T_2})'=0,~(\frac{T_0}{T_3})'\neq 0,\\
(\frac{T_0}{T_2})'\neq 0
\end{array}
$\\ \hline 3$a_2b_2(c_3)$ & 3$a_2b_2(c_2)$ & $
\begin{array}{c}
T_0\neq 0,~T_1=0,~T_2\neq 0,~T_3\neq 0,\\
T'_0\neq 0,~(\frac{T_0}{T_2})'\neq 0,~(\frac{T_0}{T_3})'=0
\end{array}
$\\ \hline
\end{tabular}
\end{center}
\begin{center}
\begin{tabular}{|l|l|l|}
\hline {\bf Cases} & {\bf MCs} & {\bf Constraints}
\\ \hline 3$a_2b_2c_4d_1(e_1)$ & $4$ & $
\begin{array}{c}
T_0\neq 0,~T_1=0,~T_2\neq 0,~T_3\neq 0,\\
T'_0\neq 0,~(\frac{T_0}{T_2})'\neq 0,~(\frac{T_0}{T_3})'\neq 0,\\
(\frac{T'_0T_2}{T_0T'_2})'\neq 0,~(\frac{T'_0T_3}{T_0T'_3})'\neq
0,\\\frac{T_2}{T_3}=constant
\end{array}
$\\ \hline 3$a_2b_2c_4d_1(e_2)$ & $3$ & $
\begin{array}{c}
T_0\neq 0,~T_1=0,~T_2\neq 0,~T_3\neq 0,\\
T'_0\neq 0,~(\frac{T_0}{T_2})'\neq 0,~(\frac{T_0}{T_3})'\neq 0,\\
(\frac{T'_0T_2}{T_0T'_2})'\neq 0,~(\frac{T'_0T_3}{T_0T'_3})'\neq
0,\\\frac{T_2}{T_3}\neq constant
\end{array}
$\\ \hline 3$a_2b_2c_4(d_2)$ & $4$ & $
\begin{array}{c}
T_0\neq 0,~T_1=0,~T_2\neq 0,~T_3\neq 0,\\
T'_0\neq 0,~(\frac{T_0}{T_2})'\neq 0,~(\frac{T_0}{T_3})'\neq 0,\\
(\frac{T'_0T_2}{T_0T'_2})'\neq 0,~(\frac{T'_0T_3}{T_0T'_3})'=0
\end{array}
$\\ \hline 3$a_2b_2c_4(d_3)$ & $4$ & $
\begin{array}{c}
T_0\neq 0,~T_1=0,~T_2\neq 0,~T_3\neq 0,\\
T'_0\neq 0,~(\frac{T_0}{T_2})'\neq 0,~(\frac{T_0}{T_3})'\neq 0,\\
(\frac{T'_0T_2}{T_0T'_2})'=0,~(\frac{T'_0T_3}{T_0T'_3})'\neq 0
\end{array}
$\\ \hline 3$a_2b_2c_4(d_4)$ & $5$ & $
\begin{array}{c}
T_0\neq 0,~T_1=0,~T_2\neq 0,~T_3\neq 0,\\
T'_0\neq 0,~(\frac{T_0}{T_2})'\neq 0,~(\frac{T_0}{T_3})'\neq 0,\\
(\frac{T'_0T_2}{T_0T'_2})'=0,~(\frac{T'_0T_3}{T_0T'_3})'=0
\end{array}
$\\ \hline 3$(a_3)$ & 3$(a_1)$ & $
\begin{array}{c}
T_0\neq 0,~T_1\neq 0,~T_2=0,~T_3\neq 0
\end{array}
$\\ \hline 3$(a_4)$ & 3$(a_1)$ & $
\begin{array}{c}
T_0\neq 0,~T_1\neq 0,~T_2\neq 0,~T_3=0
\end{array}
$\\ \hline
\end{tabular}
\end{center}
It is seen from the above tables that each case has different
constraints on the energy-momentum tensor. If we solve these
constraints, we may have exact solution of EFEs. We have attempted
(section 5) 10 different constraints to obtain the energy-momentum
tensor and the corresponding spacetime. These cases are given by
$1a_4b_1c_2d_1e_2(f_1),~2a_1b_1c_2(d_1),~2a_1b_1c_2d_2(e_1)$,
$2a_1b_2(c_1),~2a_1b_2(c_2),~2a_2b_3c_1(d_1),~2a_2b_3c_2(d_1),~
2a_2b_3c_2(d_2),~2a_2b_3c_1d_3(e_1), \\3a_2b_2c_4(d_4)$. It turns
out that the cases $2a_1b_2(c_1),~2a_1b_2(c_2),~2a_2b_3c_2(d_1),\\
2a_2b_3c_2(d_2),~2a_2b_3c_1d_3(e_1)$ provide no proper MC.
However, the cases $2a_2b_3c_1(d_1)$ and $3a_2b_2c_4(d_4)$ yield
one proper MC, the case $2a_1b_1c_2d_2(e_1)$ gives 2 proper MCs,
the case $1a_4b_1c_2d_1e_2(f_1)$ yields 3 proper MCs and the case
$2a_1b_1c_2(d_1)$ gives 4 proper MCs. It is interesting to note
that $3a_2b_2c_4(d_4)$ is the case where $T_1=0$, i.e, the
degenerate case but this provides finite dimensional MCs and we
obtain 1 proper MC in this case.

We have attempted some of the constraints to obtain exact
solutions of EFEs which give finite dimensional MCs. We have
discussed some of the physical properties of the resulting
spacetimes. It would be interesting to look for more solutions of
the constraints or examples should be constructed to satisfy the
constraints.

\newpage
\renewcommand{\theequation}{A\arabic{equation}}
\setcounter{equation}{0}
\section*{Appendix A}

The surviving components of the Ricci tensor are
\begin{eqnarray}
R_{00}&=&\frac{1}{4}e^\nu(2\nu''+\nu'^2+\nu'\lambda'+\nu'\mu'),
\nonumber\\
R_{11}&=&-\frac{1}{4}(2\nu''+2\lambda''+2\mu''+\nu'^2 +\lambda'^2
+\mu'^2),\nonumber\\
R_{22}&=&-\frac{1}{4}e^\lambda(2\lambda''+\nu'\lambda'+\lambda'^2
+\lambda'\mu'),\nonumber \\
R_{33}&=&-\frac{1}{4}e^\mu(2\mu''+\nu'\mu'+\lambda'\mu'+\mu'^2).
\end{eqnarray}
The Ricci scalar is given by
\begin{eqnarray}
R&=&\frac{1}{2}(2\nu''+2\lambda''+2\mu''+\nu'^2+\lambda'^2
+\mu'^2+\nu'\lambda'+\nu'\mu'+\lambda'\mu').
\end{eqnarray}
Using Einstein field equations (1), the non-vanishing components
of energy-momentum tensor $T_{ab}$ are
\begin{eqnarray}
T_{00}&=&-\frac{1}{4}e^\nu(2\lambda''+2\mu''+\lambda'^2
+\mu'^2+\lambda'\mu'),\nonumber\\
T_{11}&=&\frac{1}{4}(\nu'\lambda'+\nu'\mu'+\lambda'\mu'),\nonumber\\
T_{22}&=&\frac{1}{4}e^\lambda(2\nu''+2\mu''+\nu'^2
+\mu'^2+\nu'\mu'),\nonumber\\
T_{33}&=&\frac{1}{4}e^\mu(2\nu''+2\lambda''+\nu'^2
+\lambda'^2+\nu'\lambda').
\end{eqnarray}

\renewcommand{\theequation}{B\arabic{equation}}
\setcounter{equation}{0}
\section*{Appendix B}

Linearly independent KVs associated with the static cylindrical
symmetric spacetimes are given by [12]
\begin{eqnarray}
\xi_{(1)}&=&\partial_t,\nonumber\\
\xi_{(2)}&=&\partial_\theta,\nonumber\\
\xi_{(3)}&=&\partial_z.
\end{eqnarray}
The components of the energy-momentum tensor for the metric in
Eq.(70) are
\begin{eqnarray}
T_0&=&(r/r_0)^{2a}(b+c-b^2-c^2-bc)/r^2,\nonumber\\
T_1&=&(ab+bc+ca)/r^2,\nonumber\\
T_2&=&-(r/r_0)^{2b}(a+c-a^2-c^2-ac)/r^2,\nonumber\\
T_3&=&-(r/r_0)^{2c}(a+b-a^2-b^2-ab)/r^2.
\end{eqnarray}
\newpage



{\bf \large References}

\begin{description}

\item{[1]} Hall, G.S. and Costa, J.da: J. Math. Phys.
{\bf 29}(1988)2465.

\item{[2]} Hall, G.S. and Low, D.J and Pulham, J.R.: J. Math. Phys.
{\bf 35}(1994)5930.

\item{[3]} Hall, G.S.and Lonie, D.P.: Class. Quant. Gravity
{\bf 12}(1995)1007.

\item{[4]} Hall, G.S. and Costa, J.da: J. Math. Phys.
{\bf 32}(1991)2848; {\bf 32}(1991)2854.

\item{[5]} Hall, G.S., Roy, I. and Vaz, L.R.: Gen. Rel and Grav.
{\bf 28}(1996)299.

\item{[6]} Katzin, G.H., Levine J. and Davis, W.R.: J. Math. Phys.
{\bf 10}(1969)617;\\ J. Maths. Phys. {\bf 11}(1970)1578.

\item{[7]} Katzin, G.H. and Levine, J.: Tensor (NS) {\bf
22}(1971)64;\\ Colloq. Math. {\bf 26}(1972)21.

\item{[8]} Szekeres, P.: Commun. Math. Phys. {\bf 41}(1975)55.

\item{[9]} Davis, W.R. and Katzin, G.H.: Am. J. Math. Phys. {\bf
30}(1962)750.

\item{[10]} Petrov, A.Z.: {\it Einstein Spaces} (Pergamon, Oxford
University Press, 1969).

\item{[11]} Misner, C.W., Thorne, K.S. and Wheeler, J.A.: {\it
Gravitation} (W.H. Freeman, San Francisco, 1973).

\item{[12]} Kramer, D., Stephani, H., MacCallum, M.A.H. and
Hearlt, E.: {\it Exact Solutions of Einstein's Field Equations}
(Cambridge University Press, 2003).

\item{[13]} Coley, A.A. and Tupper, O.J.: J. Math. Phys. {\bf
30}(1989)2616.

\item{[14]} Hall, G.S., Roy, I. and Vaz, L.R.: Gen. Rel and Grav.
{\bf 28}(1996)299.

\item{[15]} Camc{\i}, U. and Barnes, A.: Class. Quant. Grav. {\bf
19}(2002)393.

\item{[16]} Carot, J. and da Costa, J.: {\it Procs. of the 6th
Canadian Conf. on General Relativity and Relativistic
Astrophysics}, Fields Inst. Commun. 15, Amer. Math. Soc. WC
Providence, RI(1997)179.

\item{[17]} Carot, J., da Costa, J. and Vaz, E.G.L.R.: J. Math.
Phys. {\bf 35}(1994)4832.

\item{[18]} Sharif, M.: Nuovo Cimento {\bf B116}(2001)673;\\
Astrophys. Space Sci. {\bf 278}(2001)447.

\item{[19]} Camc{\i}, U. and Sharif, M.: Gen Rel. and Grav. {\bf
35}(2003)97.

\item{[20]} Camc{\i}, U. and Sharif, M.: Class. Quant. Grav.
{\bf 20}(2003)2169-2179.

\item{[21]} Tsamparlis, M. and Apostolopoulos, P.S.: Gen. Rel.
and Grav. {\bf 36}(2004)47.

\item{[22]} Sharif, M. and Sehar Aziz: Gen Rel. and Grav. {\bf
35}(2003)1091;\\
Sharif, M.: J. Math. Phys. {\bf 44}(2003)5142.

\item{[23]} Sharif, M.: J. Math. Phys. {\bf 45}(2004).

\item{[24]} Hall, G.S.: Class. Quant. Grav. {\bf 20}(2003)4067.

\end{description}

\end{document}